\documentclass{aa}  
\usepackage{graphicx}
\usepackage{txfonts}
\usepackage{enumitem}
\usepackage{float}
\begin{document}

   \title{The diversity of spectral shapes of hydrogen Lyman lines and \ion{Mg}{ii} lines in 
                  a quiescent prominence} 
   \titlerunning{Shapes of hydrogen Lyman and \ion{Mg}{ii} line profiles}
   \author{P. Schwartz
          \inst{1}
          \and
          S. Gun\'{a}r\inst{2}
          \and
          J. Koza\inst{1}
          \and
          P. Heinzel\inst{2,3}
          }
   
   \institute{Astronomical Institute of Slovak Academy of Sciences, 05960 Tatransk\'{a} Lomnica, Slovak Republic\\
   \email{pschwartz@astro.sk}
   \and
   Astronomical Institute, Czech Academy of Sciences, 25165 Ond\v{r}ejov, Czech Republic 
   \and
  Center of Excellence - Solar and Stellar Activity, University of Wroclaw, 
  Kopernika 11, 51\,622 Wroclaw, Poland
   }

   \date{Received XXXX XX, XXXX; accepted XXXX XX, XXXX}

 
  \abstract
      {Broad sets of spectroscopic observations comprising multiple lines represent an excellent 
      opportunity for diagnostics of the properties of the prominence plasma and the dynamics of 
      their fine structures. However, they also bring significant challenges when they are compared 
      with synthetic spectra provided by radiative transfer modeling.}
      {In this work, we provide a statistical spectroscopic analysis of a unique dataset of coordinated 
      prominence observations in the Lyman lines (Ly$\alpha$ to Ly$\delta$) and the \ion{Mg}{ii} k and h 
      lines. The observed data were obtained by the Solar Ultraviolet Measurements of Emitted Radiation
      (SUMER) spectrograph on board of the Solar and Heliospheric Observatory (SoHO) satellite and
      the Interface Region
      Imaging Spectrograph (IRIS) on 22 October 2013. Only a few similar
      coordinated datasets of Lyman and \ion{Mg}{ii} k and h observations have ever been obtained in prominences 
      and we present here the first analysis using these two sets of spectral lines. Moreover, for the first 
      time, we assess the influence of noise on the statistical properties of the studied profile 
      characteristics.}
      {We focus on the following profile characteristics: the shape of the observed line profiles based on 
      the number of distinct peaks, the integrated line intensity, the center-to-peak 
      ratio describing the depth of the reversal of two-peaked profiles, and the asymmetry of these peaks.}
      {We show that the presence of noise has a negligible effect on the integrated intensity of all observed 
      lines, but it significantly affects the classification of spectral profiles using the number of distinct peaks, 
      the reversal depth,
      and also the peak asymmetry. We also demonstrate that by taking the influence
      of noise into account, we can assess which profile characteristics in which spectral lines are suitable 
      for diagnostics of different properties of the observed prominence. For example, we show that the 
      subordinate peaks (peaks below error bars) in the Lyman line profiles are mostly caused by noise, 
      which means that only the 
      dominant peaks should be used for statistical analyses or comparisons with synthetic spectra. On the 
      other hand, in the \ion{Mg}{ii} k and h profiles, the key role in the distinction between the 
      multi\discretionary{-}{-}{-}peaked profiles with low peaks and the profiles with deep reversals is played
      by the dynamics of multiple fine structures located along a line of sight. The complex, multi\discretionary{-}{-}{-}peaked
      profiles are observed in places where multiple fine structures with different line-of-sight velocities 
      are crossing the line of sight, while the profiles with deep reversals likely correspond to instances 
      when we observe single fine structures or more fine structures but with similar line-of-sight 
      velocities.}
      {This study allows us to conclude that if we are interested in the diagnostics of the dynamics of 
       prominence fine structures, the best approach is to use a combination of profile asymmetry in the 
       Lyman lines together with the complex profiles of \ion{Mg}{ii} k and h lines. On the other hand, 
       if we want to diagnose the temperature and pressure properties of individual prominence fine 
       structures, we need to focus on the deeply reversed \ion{Mg}{ii} k and h lines in combination with 
       the Lyman lines and to analyze the depth of the central reversal and the integrated intensities.}

   \maketitle
%
%
\section{Introduction}
\label{Sec:intro}
Prominences are one of the most common phenomena connected with the solar magnetic fields present in the solar 
atmosphere. They are relatively dense and cool structures supported against the gravity by coronal magnetic 
fields occurring in the low\discretionary{-}{-}{-}density and hot corona. Concise information on the 
properties of the prominence plasma and magnetic field and on the prominence modeling can be found in the reviews 
of \citet{2010SSRv..151..243L}, \citet{2010SSRv..151..333M}, 
\citet{2014IAUS..300...59G}, \citet{2014LRSP...11....1P}, and \citet{2018LRSP...15....7G}. For further 
details, readers can refer to the proceedings of the IAU 300 Symposium edited by
\citet*{2014IAUS..300.....S} or the book Solar Prominences edited by \citet{2015ASSL..415.....V},
for example.

Prominences have been observed for more than a hundred years with ground\discretionary{-}{-}{-}based instruments 
in the relatively cool chromospheric spectral lines of hydrogen, helium, and calcium in visual and near\discretionary{-}{-}{-}infrared parts of the solar spectrum. Prominences are seen in emission
above the solar limb in these lines.
On the other hand, they 
demonstrate themselves as dark filaments visible in absorption against the solar disk. Extensive progress in 
solar observations from space made it possible to obtain high\discretionary{-}{-}{-}quality ultraviolet (UV)
and extra\discretionary{-}{-}{-}ultraviolet (EUV) spectra of prominences. Such
data provided new possibilities for spectroscopic investigations of the properties of their plasma.
Spectral observations in the hydrogen Lyman line series plus continuum carried out with the  Solar
Ultraviolet Measurements of Emitted Radiation (SUMER) spectrograph
(\citeauthor{1995SoPh..162..189W} \citeyear{1995SoPh..162..189W}) on board the Solar and
Heliospheric Observatory (SoHO) represent quite substantial observational material suitable for 
analyzing the structure and physical properties of quiescent prominences using 
modelling under deviation from the local thermodynamical equilibrium (non\discretionary{-}{-}{-}LTE).

The Interface Region Imaging Spectrograph (IRIS, \citeauthor{2014SoPh..289.2733D} \citeyear{2014SoPh..289.2733D}) 
is capable of obtaining spectra with almost two times higher spatial resolution than SUMER. Moreover, the IRIS 
spectrograph can record UV spectra simultaneously with slit\discretionary{-}{-}{-}jaw images both in the
far\discretionary{-}{-}{-}ultraviolet (FUV) and near\discretionary{-}{-}{-}ultraviolet
(NUV) regions. Several UV lines (mainly \ion{Mg}{ii} k at $2796$\,\AA,
\ion{Mg}{ii} h at $2803$\,\AA, and \ion{C}{ii} $1334$\,\AA\ and $1336$\,\AA) recorded in
the IRIS spectra are suitable
for studying prominences in great detail and, in combination with modeling using the
non\discretionary{-}{-}{-}LTE radiative transfer physical properties of prominence, plasma can be derived.

In this work we present a complex statistical spectroscopic analysis of a unique dataset of coordinated 
prominence observations in the Lyman lines of hydrogen and \ion{Mg}{ii} k and h lines obtained on
22 October 2013. The observed data were recorded
quasi co\discretionary{-}{-}{-}spatially and almost simultaneously
by SUMER and IRIS. This is the first analysis of high\discretionary{-}{-}{-}resolution spectroscopic 
observations of a prominence in these two sets of spectral lines. Only a few similar coordinated datasets 
of Lyman and \ion{Mg}{ii} k and h observations have ever been obtained in prominences. The first joined 
prominence observations in the Ly$\alpha$, Ly$\beta$, \ion{Mg}{ii} k and h, and \ion{Ca}{ii} UV lines 
were carried out by the OSO\discretionary{-}{-}{-}8/LPSP 
spectrograph and  analyzed by \citet{1982ApJ...253..330V, 1982ApJ...254..780V}. 
Except for the SUMER\discretionary{-}{-}{-}IRIS 
observations that took place on 22 October 2013 and are used here, the same prominence was also
observed the next day, but
obtained spectra were strongly affected by impacts of particles from flaring active regions.
Later, another SUMER\discretionary{-}{-}{-}IRIS observing campaign for filaments and prominences 
took place in 2017 from 28 March through 4 April, during which several quiescent prominences and
filaments were observed. However, only the Ly$\alpha$ line was observed with the SUMER spectrograph 
during this campaign \citep{2021cosp...43E1770Z}. 

Broad datasets of spectroscopic observations in multiple lines represent an excellent opportunity
for diagnostics of the properties of the prominence plasma and the dynamics
of their fine structures. This was demonstrated by \citet{2010A&A...514A..43G}. These
authors developed a statistical method suitable for comparison of large datasets of
spectroscopic observations in different lines with broad sets of synthetic spectra
produced by multi\discretionary{-}{-}{}dimensional radiative transfer models. This
method was later used by \citet{2015A&A...577A..92S} to analyze SUMER observations in
the Lyman line series, with the aim to investigate the temperature and density of the
plasma forming the cores of prominence fine structures. Another example is the use of
coordinated datasets of Lyman lines observed by SUMER and H$\alpha$ observations from 
ground\discretionary{-}{-}{-}based instruments by \citet{2012A&A...543A..93G} to analyze
the dynamics of the prominence fine structures. Similarly, coordinated \ion{Mg}{ii} k and
h observations by IRIS, together with other space\discretionary{-}{-}{-}borne and
ground\discretionary{-}{-}{-}based data were used to investigate the plasma properties
and dynamics of prominences by, for instance,
\citet{2014A&A...569A..85S}, \citet{2015ApJ...800L..13H}, \citet{2016ApJ...818...31L}, and
\citet{2018ApJ...865..123R, 2019ApJ...886..134R}. 
Thanks to the availability of observations in multiple spectral lines,
\citet{2012A&A...543A..93G} and \citet{2015ApJ...800L..13H} could demonstrate that the 
H$\alpha$ line provides an important constraint on the number of the fine structures
present along line of sight (LOS). \citet{2016ApJ...818...31L} were able to analyze the
plasma and magnetic field properties of prominence legs
resembling tornado\discretionary{-}{-}{-}like vertical structures.
\citet{2018ApJ...865..123R, 2019ApJ...886..134R} could also analyze the differences in the LOS
velocities of the multiple\discretionary{-}{-}{-}component prominence fine structures.

The IRIS observations of the prominence studied here were also the focus of the work by 
\citet{2018A&A...618A..88J}. These authors analyzed the \ion{Mg}{ii} k and h, and
\ion{C}{ii} $1334$\,\AA\ and $1336$\,\AA\ spectra and compared them with
synthetic spectra produced by the 1D non\discretionary{-}{-}{-}LTE models
of \citet{2014A&A...564A.132H, 2015ApJ...800L..13H}. Using the temperature-sensitive ratio of integrated 
intensities of the \ion{Mg}{ii} k and h lines, \citet{2018A&A...618A..88J} concluded that
the observed prominence plasma might have kinetic temperature as low as $5,000$\,K.
Moreover, these authors found no evidence of global
oscillations in the studied prominence, but concluded a likely presence of random motions of fine
LOS velocities on the order of $\pm10$\,km\,s$^{-1}$. The \ion{Mg}{ii} k and h spectra of other
prominences observed by IRIS were used for analyses of the prominence dynamics and plasma properties also by 
\citet{2021A&A...653A...5P}. These authors developed a method for 
profile\discretionary{-}{-}{-}to\discretionary{-}{-}{-}profile comparison of the observed \ion{Mg}{ii} k and h 
lines with synthetic profiles produced by the 1D non\discretionary{-}{-}{-}LTE models 
of \citet{2004ApJ...617..614L}, some of which included the 
prominence\discretionary{-}{-}{-}corona transition\discretionary{-}{-}{-}region
(PCTR) extension developed by \citet{2019A&A...625A..30L}. \citet{2021A&A...653A...5P} found that when static
single\discretionary{-}{-}{-}slab models are used, a satisfactory agreement between the synthetic and observed 
profiles can be found only in the outer layers of the observed prominence, where complex \ion{Mg}{ii} k and h 
profiles are not present. These findings were confirmed by \citet{2021A&A...653A..94B}, who analyzed a different 
prominence using the method of \citet{2021A&A...653A...5P}. Another prominence simultaneously observed in the 
\ion{Mg}{ii} k and h and the H$\alpha$ line was analysed by \citet{2022ApJ...932....3J}. These authors developed 
a non\discretionary{-}{-}{-}LTE inversion method that relies on a wide grid of 1D non\discretionary{-}{-}{-}LTE 
models and fitting of five parameters of the observed spectra -- the integrated intensity of \ion{Mg}{ii} k and 
H$\alpha$, widths of both lines, and the ratio of integrated intensities of \ion{Mg}{ii} k and \ion{Mg}{ii} h 
lines. 

The current paper presents the first statistical analysis of the quasi\discretionary{-}{-}{-}simultaneous 
observations of a quiescent prominence in the \ion{Mg}{ii} k and h line and the Lyman lines including the 
Ly$\alpha$ line. Moreover, for the first time, we take into account the influence of the noise on the 
following profile characteristics: the shape of the observed line profiles based on the number of 
distinct peaks, the integrated intensity, the depth of the central reversal of
two\discretionary{-}{-}{-}peaked profiles, and the asymmetry of their peaks. 
%
\section{Observations}
\label{Sec:observations}
\begin{figure*}
\centering
\resizebox{0.98\hsize}{!}{
\includegraphics{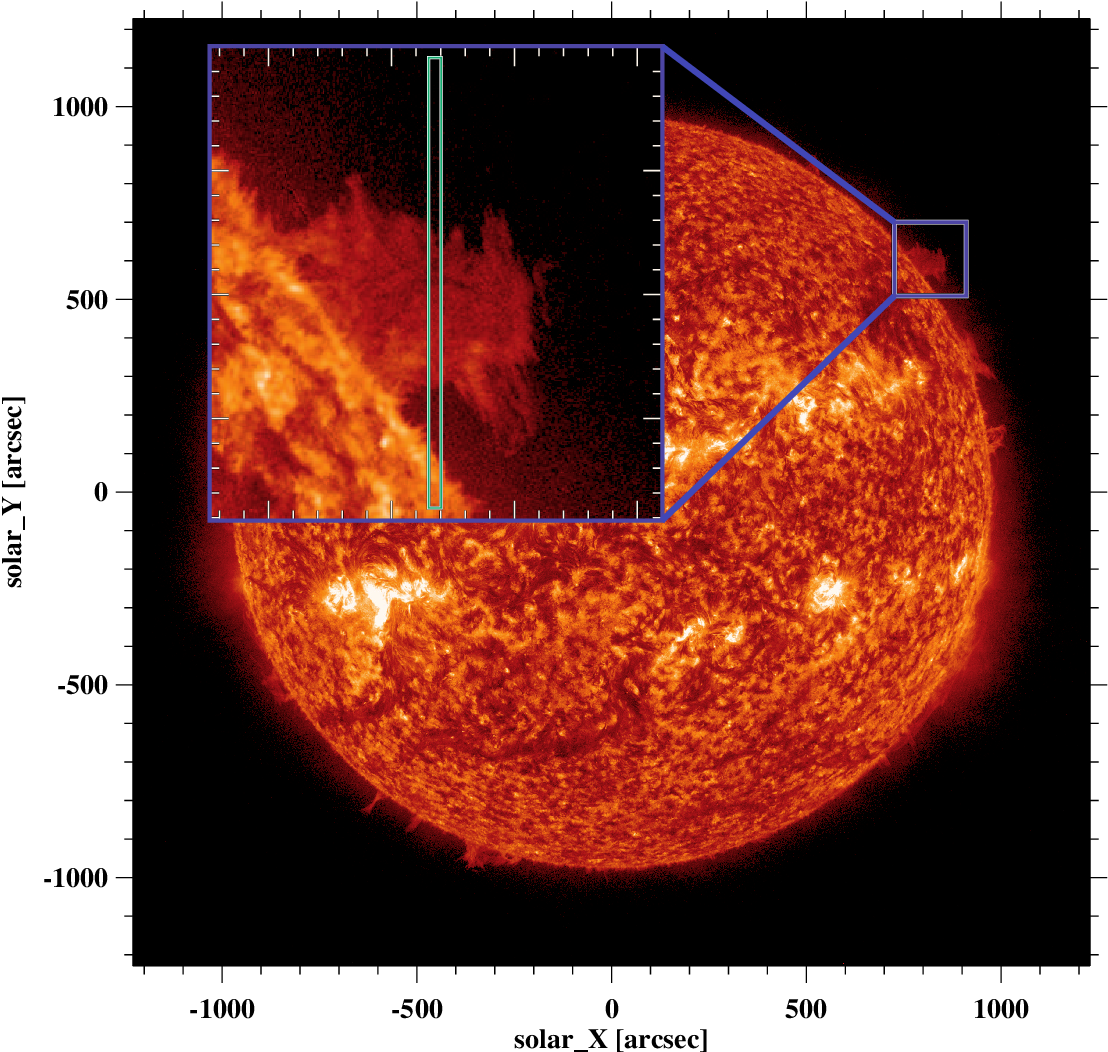}}
\caption{Position and a general shape of the observed prominence at the limb shown in the 
$304$\,\AA\ AIA channel. The zoomed image of the prominence reveals its basic structure. The position
of the IRIS raster is indicated by the green rectangle.}
\label{Fig:BasicObsProm}
\end{figure*}
\begin{figure*}
\parbox{0.5094\hsize}{ 
\resizebox{\hsize}{!}{
\includegraphics{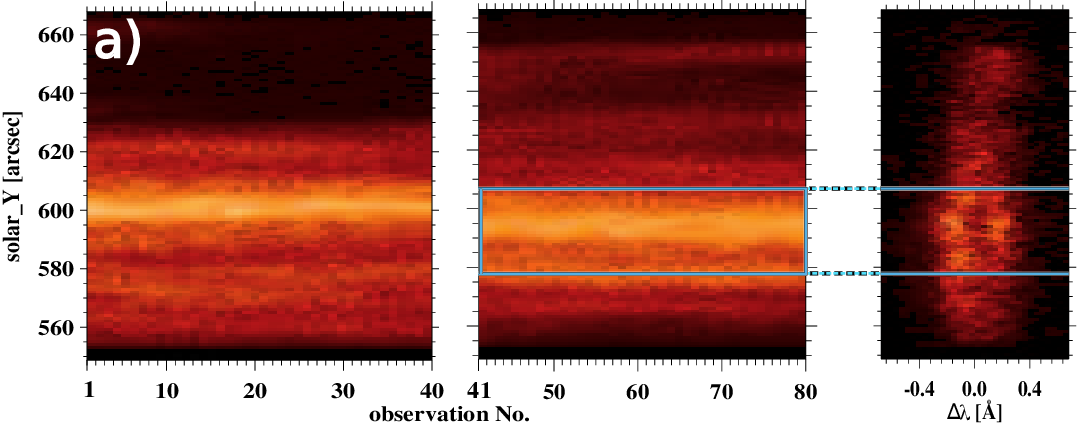}}}\
\parbox{0.02\hsize}{\phantom{xx}}
\parbox{0.4605\hsize}{ 
\resizebox{\hsize}{!}{
\includegraphics{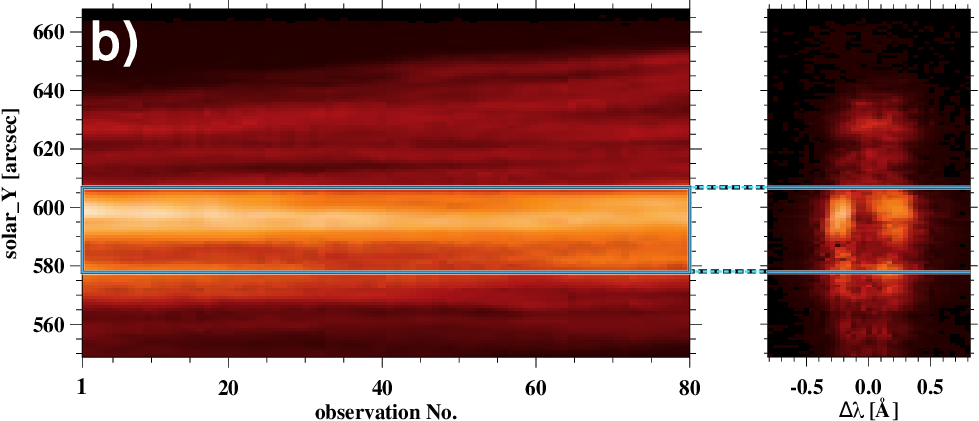}}}
\caption{Maps of integrated intensities of the Ly$\beta$ (\textsf{a}) and Ly$\alpha$ (\textsf{b}) S\hspace{0.1ex}\&\hspace{0.1ex}S observations of the SUMER spectrograph
(left part of each panel). The abscissas show the observation numbers counted from
the beginning of the observations of a given spectral line (see observation No.\,$1$ in
Table~\ref{tab:SumObs}). In panel \textsf{a}, both Ly$\beta$ blocks of
observations are put together, even though they did not follow one after the other,
to show that the pointing of the instrument differs between them. The right part of each
panel shows an example of the observed spectra of the Ly$\beta$ and Ly$\alpha$ lines;
the relative wavelength scale according to the laboratory central wavelength
(Table~\ref{tab:LineList}) is indicated. The horizontal blue lines mark the part along
the slit from which the data were taken for the analysis. As for Ly$\beta$, these lines
occur only in the latter block of the Ly$\beta$ observations (the block of
No.\,5 observations, see Table~\ref{tab:SumObs}) because the pointing was not correct during
the first block of observations.}
\label{Fig:SUMERobs}
\end{figure*}
\renewcommand{\arraystretch}{1.1}
\begin{table}
\caption{Basic information on spectral lines observed by SoHO/SUMER and IRIS spectrographs. }
\label{tab:LineList}
\centering 
\begin{tabular}{ccccc}
\hline\hline\\[-1.8ex]
line & ion & $\lambda_{\mathrm{0}}\,\left[\AA\right]$ & transition & instr.\\[0.6ex]
\hline\\[-1.7ex]
Ly$\alpha$ & \ion{H}{i} & $1215.67$ & 2p~$^2\mathrm{P}_{3/2,\,1/2}$\,---\,1s~$^2\mathrm{S}_{1/2}$ & 
   $\mathcal{S}$ \\[1.01ex]
Ly$\beta$  & \ion{H}{i} & $1025.72$ & \hspace{3.0ex}3p~$^2\mathrm{P}_{3/2}$\,---\,1s~$^2\mathrm{S}_{1/2}$ & 
   $\mathcal{S}$ \\[1.01ex]
Ly$\gamma$ & \ion{H}{i} & \hspace{1ex}$972.54$ & \hspace{3.0ex}4p~$^2\mathrm{S}_{1/2}$\,---\,1s~$^2\mathrm{S}_{1/2}$ & 
   $\mathcal{S}$ \\[1.01ex]
Ly$\delta$ & \ion{H}{i} & \hspace{1ex}$949.74$ & \hspace{3.0ex}5p~$^2P_{3/2}$\,---\,1s~$^2\mathrm{S}_{1/2}$ & 
   $\mathcal{S}$ \\[1.01ex]
\ion{Mg}{ii}~k & \ion{Mg}{ii} & $2796.35$ & \hspace{3.0ex}3p~$^2\mathrm{P}_{3/2}$\,---\,3s~$^2\mathrm{S}_{1/2}$ & 
  $\mathcal{I}$ \\[1.01ex]
\ion{Mg}{ii}~h & \ion{Mg}{ii} & $2803.53$ & \hspace{3.0ex}3p~$^2\mathrm{P}_{1/2}$\,---\,3s~$^2\mathrm{S}_{1/2}$ & 
$\mathcal{I}$ \\
\hline
\end{tabular}
\tablefoot{
$\lambda_{\mathrm{0}}$ -- laboratory central wavelengths for vacuum \\
\phantom{xxx}\hspace{0.9ex}instr. -- instrument; $\mathcal{S}$ for SoHO/SUMER, $\mathcal{I}$ for IRIS
}
\end{table}
%
%
\renewcommand{\arraystretch}{1.1}
\begin{table*}
\caption{Summary of the SUMER observations of the Ly$\alpha$\,--\,Ly$\delta$ lines.}
\label{tab:SumObs}
\centering 
\begin{tabular}{ccccccc}
\hline\hline 
block of obs.  &  spectral  & observa-   & time in UT           & exposure    & cadence   &  pointing  \\    
 No.           &  line      & tions No.  & beginning\,--\,end    & time [s]     &  [s]      &       
 setting   \\
\hline
1              & Ly$\beta$  & \hspace{1ex}$1$\,--\,$40$     & 07:00:52\,--\,07:07:22   & $10$     & $10$  & preliminary \\
2              & Ly$\gamma$ & \hspace{1ex}$1$\,--\,$40$     & 07:07:41\,--\,07:20:41   & $20$     & $20$  & preliminary \\
3              & Ly$\delta$ & \hspace{1ex}$1$\,--\,$40$     & 07:21:10\,--\,07:34:10   & $20$     & $20$  & preliminary \\
4              & Ly$\alpha$ & \hspace{1ex}$1$\,--\,$80$     & 07:45:54\,--\,07:59:05   & $10$     & $10$  & final  \\
5              & Ly$\beta$  & $41$\,--\,$80$                & 07:59:44\,--\,08:06:14   & $10$     & $10$  & final \\
6              & Ly$\gamma$ & $41$\,--\,$80$                & 08:06:34\,--\,08:19:34   & $20$     & $20$  & final \\
7              & Ly$\delta$ & $41$\,--\,$80$                & 08:20:02\,--\,08:33:02   & $20$     & $20$  & final \\
\hline
\end{tabular}
\end{table*}
A quiescent prominence located at the NW limb was observed on 22 October 2013 
quasi\discretionary{-}{-}{-}simultaneously  
and almost co\discretionary{-}{-}{-}spatially with the SUMER and IRIS 
space\discretionary{-}{-}{-}born spectrographs in the Lyman series of hydrogen and in
the \ion{Mg}{ii} k and h lines, respectively. Basic information on the spectral lines
observed by the two spectrographs are listed in Table~\ref{tab:LineList}.
Position of the observed prominence at the limb is shown in the
full\discretionary{-}{-}{-}disk image in Fig.~\ref{Fig:BasicObsProm} recorded in the
$304$\,\AA\ channel of the Atmospheric Imaging Assembly
(AIA; \citeauthor{2012SoPh..275...17L} \citeyear{2012SoPh..275...17L}) on board the Solar
Dynamics Observatory
(SDO; \citeauthor{2012SoPh..275....3P} \citeyear{2012SoPh..275....3P}).
The helioprojective\discretionary{-}{-}{-}Cartesian coordinate system (with abscissa
denoted as solar\_X and ordinate as solar\_\kern-0.7ptY) is used in this image and hereafter in
text and images of this work. In the zoomed image, the field of view (IRIS FOV) of
the IRIS spectrograph during its raster observations is marked. Position of IRIS FOV
is fixed according to an information given in the data files before any
co\discretionary{-}{-}{-}alignment was carried out. Because pointing of SUMER during
the observations, as given in the data files, is not reliable, the slit position of
SUMER can be estimated by co\discretionary{-}{-}{-}alignment with IRIS.

It is important also to note that observed spectral profiles are taken as they are 
(the level\discretionary{-}{-}{-}2 data after basic reduction and radiometric calibration) 
without applying additional corrections for instrumental effects as deconvolution of 
instrumental profile, correction for point\discretionary{-}{-}{-}spread function, etc. 
Because methods used for correction for these effects are mathematically complicated and 
unstable, influence of these effects will be added to synthetic profiles calculated 
by the non\discretionary{-}{-}{-}LTE modeling before comparison with results of analysis
of observed data presented here. This will be done in our next paper.
\subsection{Lyman line observations}
\label{Sec:sumer}
SUMER observations were made between 07:45 and 08:33\,UT with a fixed slit position -- the 
so\discretionary{-}{-}{-}called sit\discretionary{-}{-}{-}and\discretionary{-}{-}{-}stare 
observation mode (hereafter referred to as S\hspace{0.1ex}\&\hspace{0.1ex}S observations) -- 
pointed at $\mathrm{solar\_X}=799.94\,\mathrm{arcsec}$ and
$\mathrm{solar\_}\kern-0.7pt\mathrm{Y}=630.0\,\mathrm{arcsec}$
according to headers of the data files. The height and width of the slit were 
$119.6$\,arcsec and $0.28$\,arcsec, respectively. Size of one pixel along the slit corresponds 
to $1$\,arcsec in the helioprojective \hbox{solar\_\kern-0.7ptY} coordinate. Due to roll of 
the satellite during the observations the slit of the spectrograph was inclined by $4.9\,\deg$ 
in counter\discretionary{-}{-}{-}clockwise direction from original N\,--\,S direction. 
The S\hspace{0.1ex}\&\hspace{0.1ex}S observations 
of the four hydrogen Lyman lines Ly$\alpha$\,--\,Ly$\delta$ were made in individual blocks, 
unlike the observations used in \citet{2010A&A...514A..43G} and \citet{2015A&A...577A..92S} 
where individual spectral lines were alternated periodically. The four Lyman lines were observed in five spectral windows: The Ly$\alpha$ spectra were observed roughly divided
into their red and blue halves placed on two spectral windows observed simultaneously
occurring side\discretionary{-}{-}{-}by\discretionary{-}{-}{-}side on the bare part of
the spectrograph detector B, that is, both outside of attenuator and also outside the
KBr\discretionary{-}{-}{-}coated part of the detector. Calibration of data recorded
at the attenuator, which decreases radiation incident on the detector, is not reliable.
On the other hand, KBr coating amplifies the radiation, which would increase danger of
damage of the detector by this very intense line. The observed data were reduced and
calibrated using standard Solar Soft procedures for the SUMER data reduction (see
\citeauthor{SUMER_Cookbook} \citeyear{SUMER_Cookbook} and references therein).
Processing of the data was carried out using the procedures in the following order: decompression
of binary data, dead\discretionary{-}{-}{-}time correction,
flat\discretionary{-}{-}{-}fielding, local\discretionary{-}{-}{-}gain correction, and correction for the geometrical distortion of the detector. Data from the two Ly$\alpha$
spectral windows was joined together before applying correction for the geometrical distortion. Finally, the data were calibrated to radiometric units using respective Solar
Soft procedure. To extend lifetime of the detector as much as possible, a lower detector
voltage of $5412$\,V was used instead of the standard $5656$\,V for all observations. Using
this lower voltage affects calibrated Lyman lines specific intensities only marginally;
for instance, it was shown by \citet{2012SoPh..281..707S} that using of even lower voltage of
$5390$\,V changed calibrated L$\alpha$ intensities by less than $2$\,\%.
To better ensure that the detector does not get overburned
when not using the attenuator, the
aperture door of the instrument was partially closed during the Ly$\alpha$ line
observations. Therefore, the specific intensities of this line were multiplied by a
geometric factor of five after the radiometric calibration. This factor comes from
ratio of area of the instrument entrance when door is fully opened to the area with
partially closed door. The wavelength ranges of the two spectral windows were
$1213.57$\,--\,$1215.68$\,\AA\ and $1215.72$\,--\,$1217.82$\,\AA. The three higher
Lyman lines Ly$\beta$\,--\,Ly$\delta$ were observed sequentially each of them in one
spectral window on the bare part of the detector B. The wavelength ranges of the
spectral windows were $1024.67$\,--\,$1026.82$\,\AA, $971.48$\,--\,$973.64$\,\AA,
and $948.68$\,--\,$950.84$\,\AA. Size of one spectral pixel was $0.04$\,\AA\ in all
five spectral windows.

The observations in the individual Lyman lines were performed in blocks in the following sequence: 
Ly$\beta$, Ly$\gamma$, Ly$\delta$, Ly$\alpha$, Ly$\beta$, Ly$\gamma$, and Ly$\delta$. 
Due to technical problems 
with the pointing mechanism of the instrument, one cannot be sure whether position of the spectrograph
slit is correct when set automatically. Therefore the slit position was set up interactively during the 
first three blocks of observations in order to position the slit at the prominence within the IRIS 
raster as reliably as possible. Therefore, data of Ly$\beta$, Ly$\gamma$, Ly$\delta$ obtained during 
the first three blocks of observations was not used in the analysis. Then the following four blocks
of observations of Ly$\alpha$\,--\,Ly$\delta$ were made with final pointing as can be seen in 
Fig.~\ref{Fig:SUMERobs}. The figure shows the map of the Ly$\beta$ integrated intensities with number 
of observation on abscissa and position along the slit in the solar\_\kern-0.7ptY coordinates in 
arcsec (measured from the S pole) on ordinate. The two blocks of the Ly$\beta$ observations are shown 
side\discretionary{-}{-}{-}by\discretionary{-}{-}{-}side in the map in the left panel of the figure 
although they were not observed right one after another. In the map of Ly$\alpha$ integrated 
intensities in the right panel (\textsf{b}) of the figure, it can be seen that the whole Ly$\alpha$ 
observations were done with the final pointing. More information on all seven blocks of 
S\hspace{0.1ex}\&\hspace{0.1ex}S observations is listed in Table~\ref{tab:SumObs} where the blocks of 
observations made with the final pointing are indicated  with the word ``\,final\,'' in the last
column. Only data from these blocks was taken into our spectroscopic analysis. There are $80$
observations of Ly$\alpha$ made with the final pointing and $40$ such observations for each of
the Ly$\beta$\,--\,Ly$\delta$ spectral lines.

\begin{figure*}
\parbox{0.48\hsize}{
\resizebox{0.95\hsize}{!}{
\includegraphics{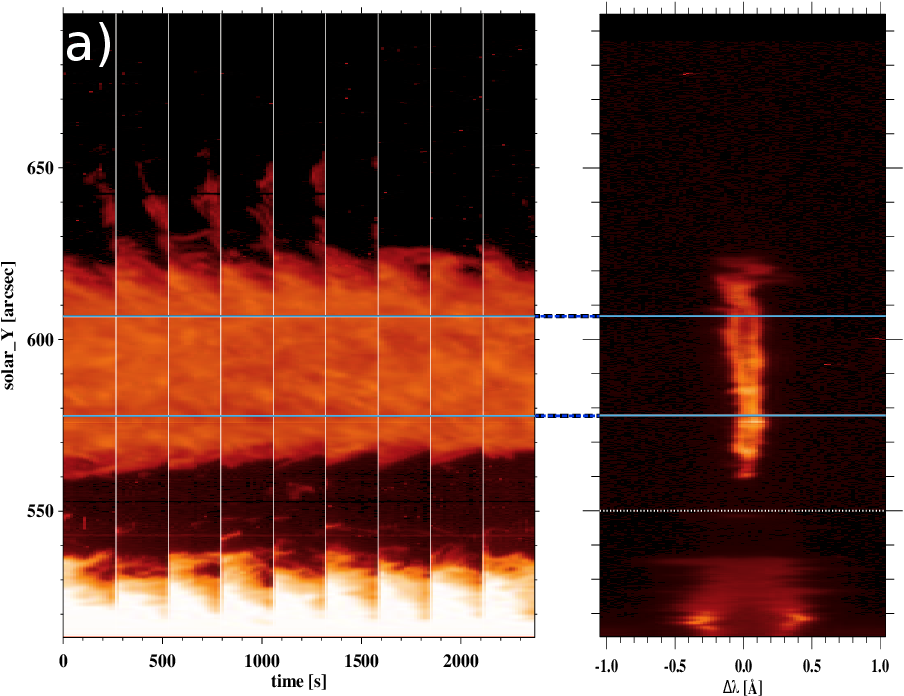}}}\
\parbox{0.02\hsize}{\phantom{1ex}}
\parbox{0.48\hsize}{
\resizebox{0.95\hsize}{!}{
\includegraphics{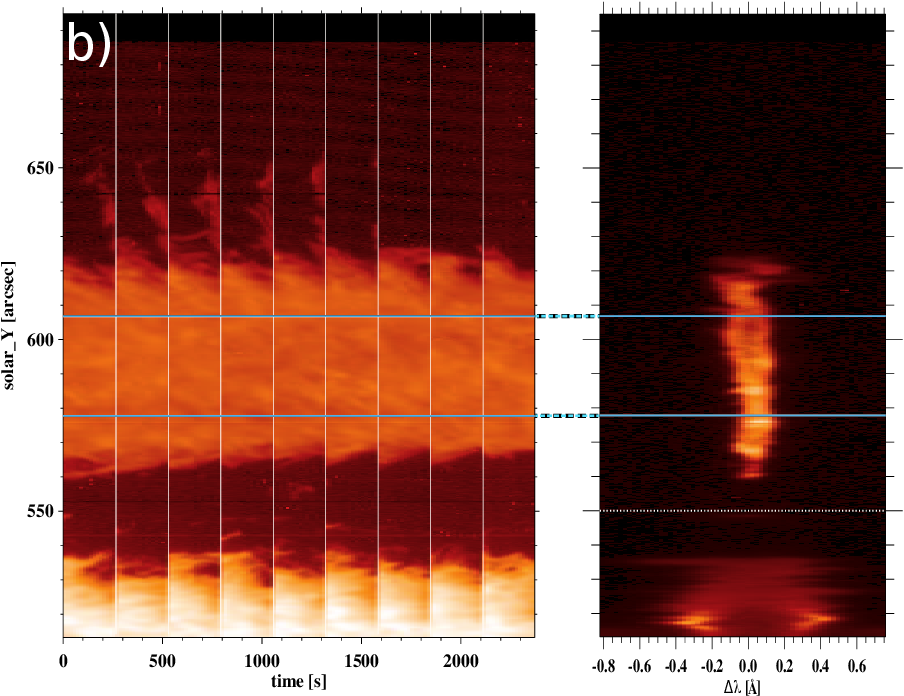}}}
\caption{Integrated intensities and examples of spectra of \ion{Mg}{ii} k (a) and \ion{Mg}{ii} h
(b) from the first nine rasters taken into account for the analysis. Time runs from
the beginning of the first raster made at 08:40:00 UT. The wavelength
scales $\Delta \lambda$ are relative to the line center wavelengths $\lambda_{\rm 0}$ 
(Table~\ref{tab:LineList}). 
The blue horizontal lines indicate the section from which line profiles were taken into 
the analysis. In order to achieve a sufficient visibility of both \ion{Mg}{ii} spectral lines 
in the spectra at the disk edge, limb and prominence, specific intensities for solar\_Y below 
$550$\,arcsec (white dotted horizontal line) were decreased by the factor of $0.3$. 
} 
\label{Fig:IRISobs}
\end{figure*}

Because direction of dispersion of spectra is not exactly parallel with a detector edge,
mutual shifts in positions along the slit between different wavelengths occur. Moreover,
effect of nonlinearity of the grating
focus\discretionary{-}{-}{-}mechanism causes additional contributions to these shifts
\citep{SUMER_Cookbook}. The shifts for the Ly$\beta$\,--\,Ly$\delta$ lines with respect
to Ly$\alpha$ were estimated by mutual comparing of distributions of their integrated
intensities along the slit. Subsequently, positions along the slit for the spectra of
the Ly$\beta$\,--\,Ly$\delta$ were corrected for these shifts. Errors of measured specific
intensities were estimated using the Poisson's statistics. Thus, relative errors of specific 
intensities are equal to reciprocal of the square root of uncalibrated intensities in counts. 
An additional error arises from uncertainty of the SoHO/SUMER radiometric calibration. This 
uncertainty was estimated to be within $15$\,\% by 
\citet{1997SoPh..170...75W} and \citet{1998ApOpt..37.2646S} during early years after the
SUMER launch. One can assume that the calibration uncertainty in the year 2013 could be
larger after almost 20 years of SUMER operation.

Assuming that the width of the instrumental profile is determined mainly by the width of the 
slit image on the detector, it was found that the instrumental profile is much narrower 
than the width of the Ly$\alpha$ profiles observed in the prominence when using such a narrow 
slit of $0.28$\,arcsec. It was already shown by \citet{2020A&A...644A.109G} that in a such  
case observed Ly$\alpha$ profile intensities are modified by deconvolution of the 
instrumental profile by up to $5$\,\%, that is below uncertainty of the radiometric calibration. 
We found that the intensities of the Ly$\beta$\,--\,Ly$\delta$ line profiles are modified 
by the deconvolution by up to $12$\,\%, that is also below the uncertainty. Thus, the 
influence of the instrumental profile on the observed profiles of all four Lyman lines 
was neglected in our further analysis. 
%
\subsection{Observations in the \ion{Mg}{ii} k and h lines}
\label{Sec:iris}
The IRIS spectrograph observed the prominence between 08:40 and 10:52\,UT shortly after the  of 
the SUMER observations. Thirty dense rasters were made by IRIS from which only first nine  
rasters made between 08:40 and 09:19\,UT were taken into our analysis due to the 
fact that later IRIS observations were affected by impacts of 
high\discretionary{-}{-}{-}energetic particles. 
These were produced by the C1.0\,--\,C1.7 class flares which occurred in the 
active region AR~11875 located close to the center of the solar disk between 09:07 and 
09:24\,UT\footnote{\url{https://www.solarmonitor.org/?date=20131022}}.  
Each of the rasters was composed of 16 slit positions 
with the step in the solar\_X direction of $0.34\,\mathrm{arcsec}$ and the 
cadence of $16.5\,\mathrm{s}$ and lasted $4$ minutes and $24$ seconds. Spectra in eight wavelength windows -- six of them encompassing the spectral lines of 
\ion{C}{ii}~$1334$\,\AA,
\ion{C}{ii}~$1336$\,\AA, \ion{Fe}{xii}~$1349$\,\AA, \ion{0}{i}~$1356$\,\AA,
\ion{Si}{iv}~$1394$\,\AA, \ion{Si}{iv} $1403$\,\AA, and \ion{Mg}{ii} k and h, and two windows
comprising wavelengths around $2832$ and $2814$\,\AA\ -- were recorded simultaneously
together with slit\discretionary{-}{-}{-}jaw images (hereafter referred to as SJ images) in 
the two channels \ion{Mg}{ii} k and \ion{Si}{iv}~$1400$\,\AA. All spectra in all eight spectral
windows and also SJ images were recorded with exposure time of $15\,\mathrm{s}$. Both \ion{Mg}{ii} 
k and h spectral lines occur in the \ion{Mg}{ii} $2796$\,\AA\ window. Spectra of the \ion{Mg}{ii}
lines were extracted from this wavelength window in the wavelength intervals 
$2795.10$\,--\,$2797.62$ and $2802.23$\,--\,$2804.77$\,\AA, respectively. 
The level\discretionary{-}{-}{-}2 IRIS data -- data after all corrections were applied
including the conversion to $\mathrm{DN}\,\mathrm{s}^{-1}$ but excluding the radiometric
calibration -- from all wavelength windows was downloaded as
multi\discretionary{-}{-}{-}extension FITS files with keywords standardly recognized by
the Solar Soft (see sect. 3.  IRIS Level 2 Data in the IRIS Technical Note ITN~No.\,26
by \citet{ITN26_ref}).

According to information stored in headers of the level\discretionary{-}{-}{-}2 IRIS FITS files, 
pointing of the center of all 30 rasters in Cartesian coordinate system was 
\mbox{$\mathrm{solar\_X}=818.17\,\mathrm{arcsec}$} and
\mbox{$\mathrm{solar\_}\kern-0.7pt\mathrm{Y}=603.89\,\mathrm{arcsec}$}
and their FOV was $5.52\,\mathrm{arcsec}\hspace{0.5ex}\times\,181.65\hspace{0.5ex}\mathrm{arcsec}$ 
(see the green rectangle in Fig.~\ref{Fig:BasicObsProm}). The size of one pixel along the slit (solar\_\kern-0.7ptY direction) is $0.166\,\mathrm{arcsec}$. The spectra and SJ images were 
taken with native spatial and spectral resolutions. Maps of integrated intensities of the k 
and h lines in the nine rasters taken into the analysis together with examples of the spectra 
are shown in Fig.~\ref{Fig:IRISobs}. 

The spectral data of both \ion{Mg}{ii} lines in DN was converted into the physical units
$\mathrm{W}\,\mathrm{m}^{-2}\mathrm{sr}^{-1}\mathrm{\AA}^{-1}$ according to the method
described in sect. 5.2 Radiometric Calibration of ITN~No.\,26 \citep{ITN26_ref} and Eq.~(1) of
\citet{2015SoPh..290.3525L}. The specific intensities were subsequently converted to
$\mathrm{erg}\,\mathrm{s}^{-1}\,\mathrm{cm}^{-2}\,\mathrm{Hz}^{-1}$,
as it is favorable for comparison with synthetic profiles calculated by our non\discretionary{-}{-}{-}LTE models. Parameters necessary for the radiometric
calibration, especially the effective area for the date of observations, were obtained
using the Solar Soft \texttt{iris\_get\_response.pro} procedure. The parameters defining
the solid angle
(see, e.g., \citeauthor{2021ApJS..255...16G} \citeyear{2021ApJS..255...16G}) 
corresponding to one pixel on the detector which also enters in the formula for the radiometric calibration, were inferred from FITS\discretionary{-}{-}{-}file headers. The precision of the 
radiometric calibration, which was determined mainly from uncertainties of the effective area estimated 
from the cross\discretionary{-}{-}{-}calibration with the SORCE/SOLSTICE instrument 
\citep{2005SoPh..230..259M}, is within $5$\,\% 
\citep{2021ApJS..255...16G}. 
Because this is below the noise in the measured specific intensities, we do not consider 
this uncertainty in our analysis.
Errors of measured specific intensities are combination of photon counting errors
and readout noise (dark current). The photon counting errors are obtained using Poisson
statistics out of registered signal. Readout noise is function of three detector parameters
cha\-rac\-te\-ri\-zing number of electrons produced by a photon impact, number of electrons required
to record one DN, and readout uncertainty \citep{2014SoPh..289.2733D}. Then, errors of measured
intensities are calculated as square root of sum of squares of the two errors. Moreover, an additional
constant readout noise of the detector is simply added to the errors of measured  intensities,
as it is shown, for example, in Eq.~(3) of \citet{2021ApJS..255...16G}.
This additional readout noise is uniform
for the whole detector and its value is $3$\,DN \citep{2015ApJ...811..127S}.
\begin{figure*}
\parbox{0.49\hsize}{
\resizebox{\hsize}{!}{
\includegraphics{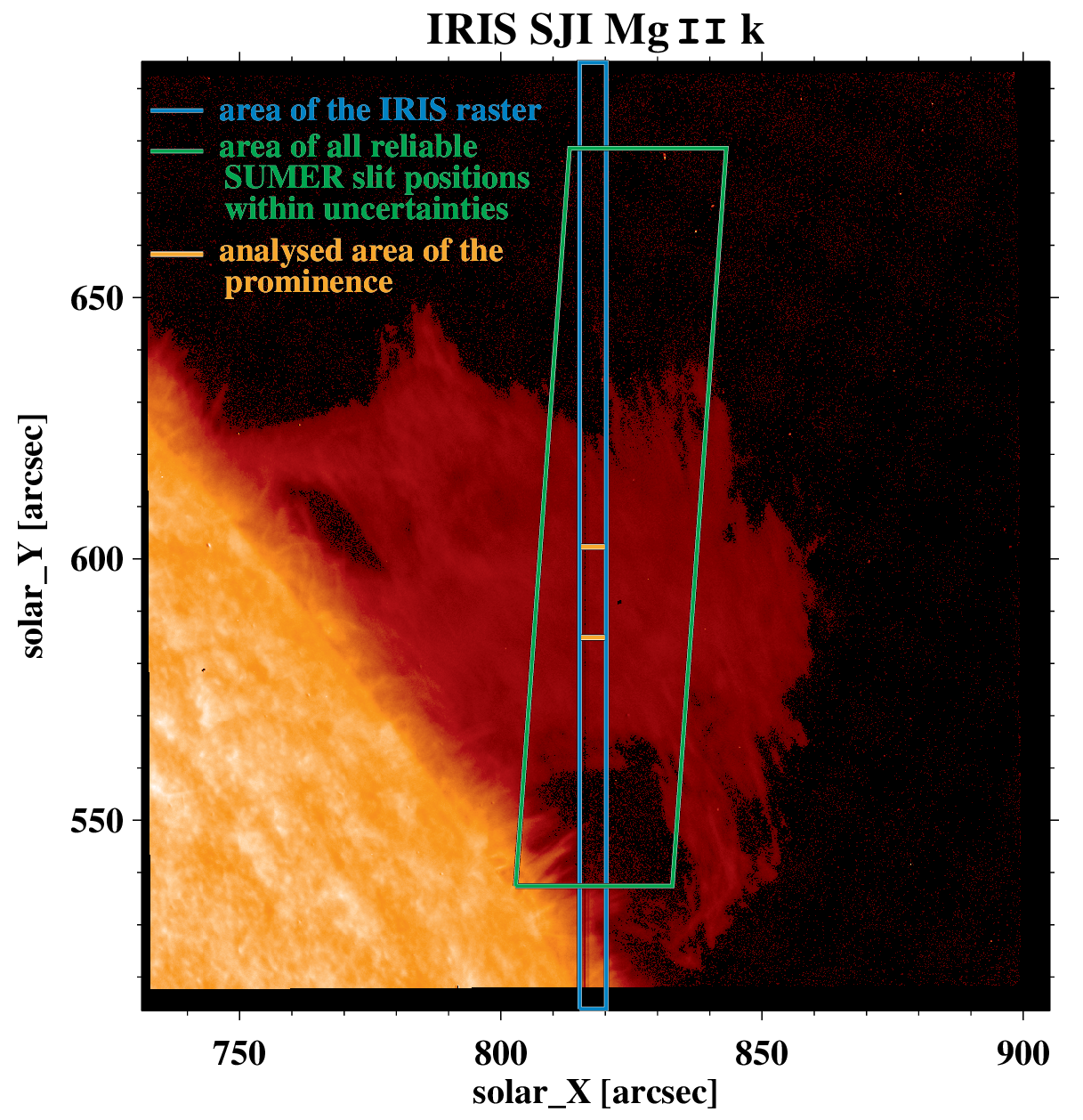}}}
\parbox{0.02\hsize}{\phantom{1ex}}
\parbox{0.49\hsize}{
\resizebox{\hsize}{!}{
\includegraphics{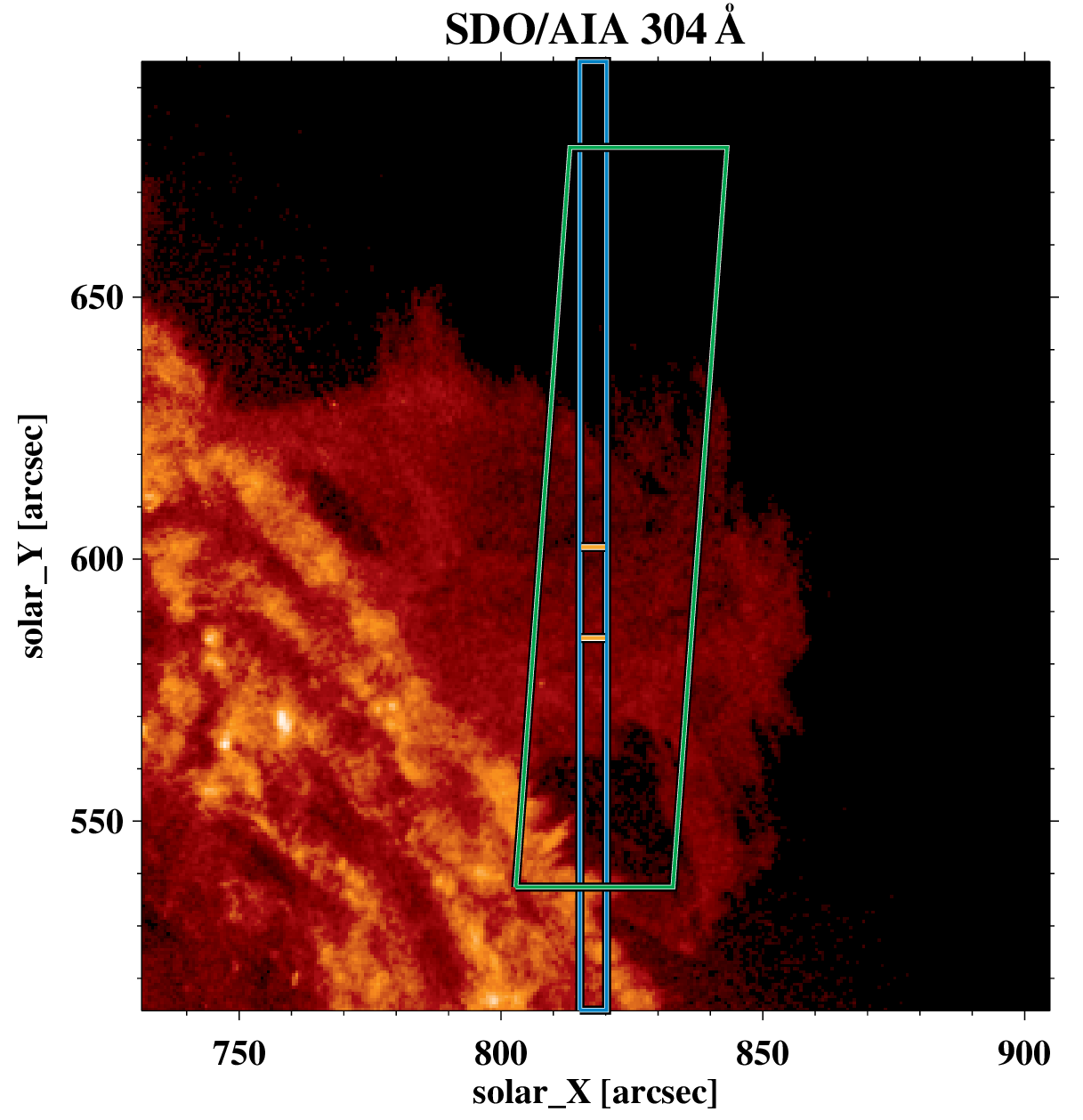}}}\\
\parbox{0.49\hsize}{
\resizebox{\hsize}{!}{
\includegraphics{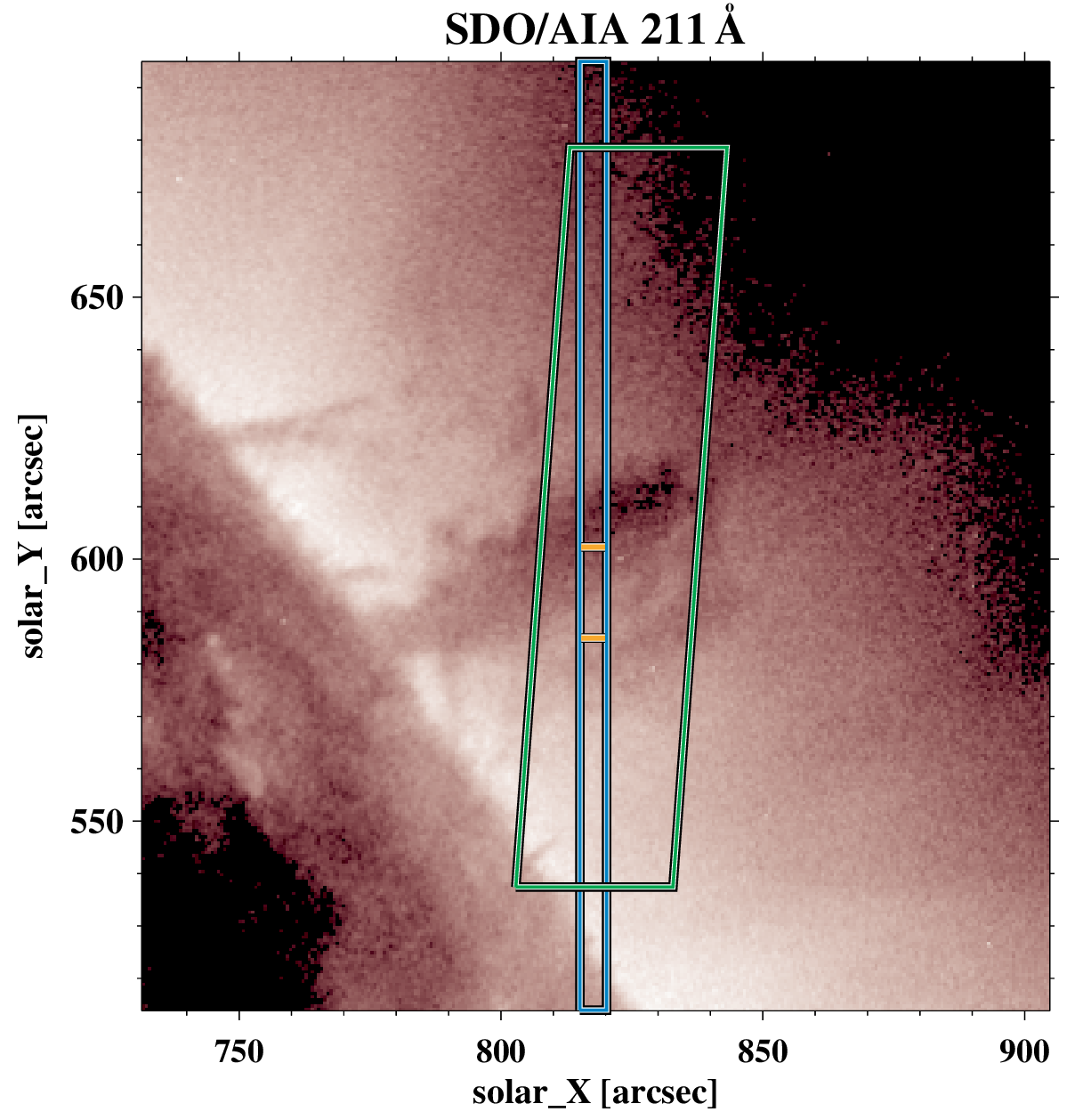}}}
\parbox{0.02\hsize}{\phantom{1ex}}
\parbox{0.49\hsize}{
\resizebox{\hsize}{!}{
\includegraphics{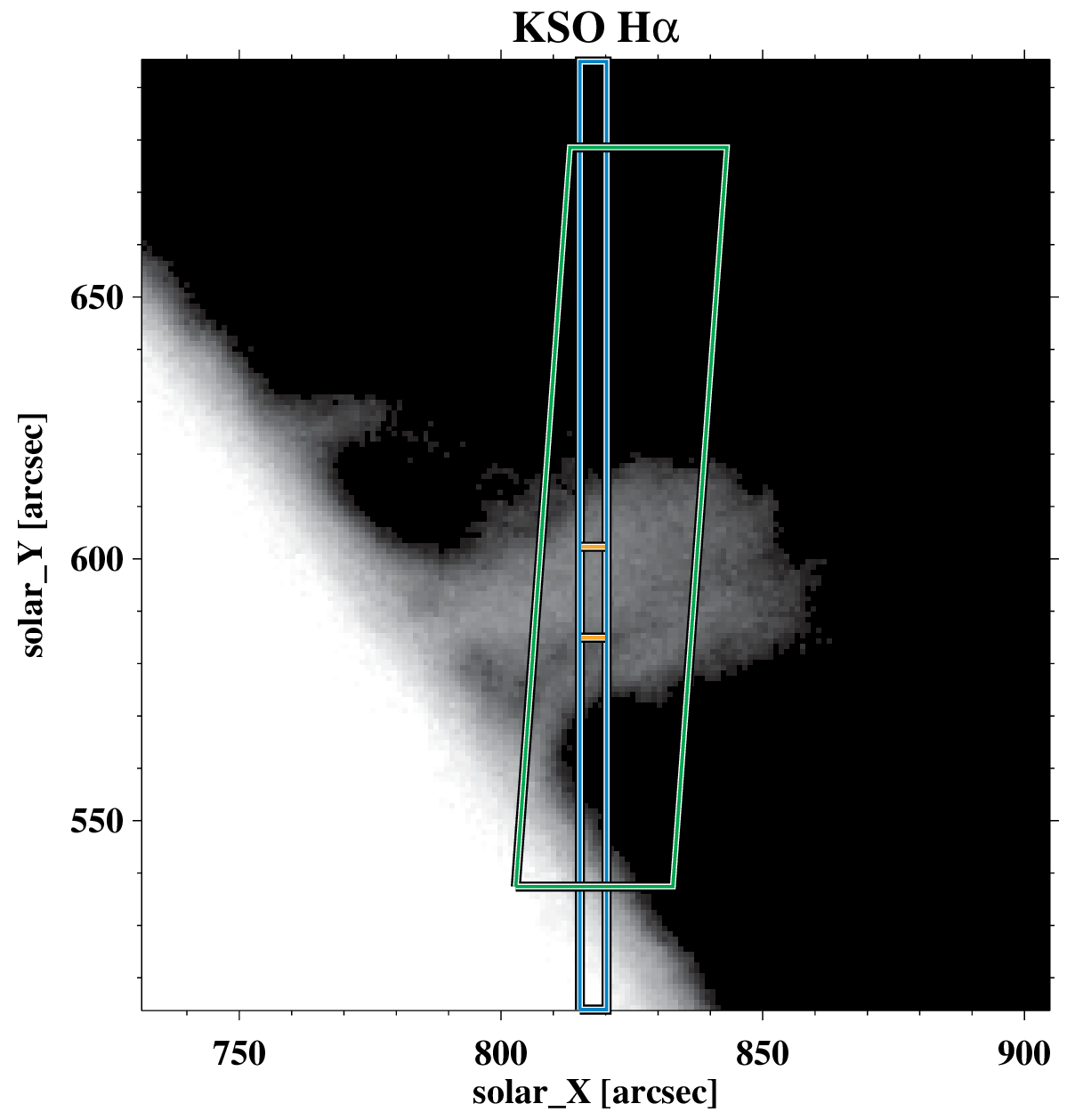}}}
\caption{ IRIS \ion{Mg}{ii} SJ image recorded at 08:40:16\,UT (upper left panel)
         co\discretionary{-}{-}{-}aligned by default with the 
         AIA $304$\,\AA\ channel image obtained at 08:40:19\,UT (upper right panel). 
         In the lower left panel is 
         the AIA $211$\,\AA\ channel image from 08:40:12\,UT and in the lower right panel is the
         KSO H$\alpha$ filtergram recorded at 07:32:01\,UT;  
         both these images are co\discretionary{-}{-}{-}aligned 
         with the IRIS observations. Structures visible in the H$\alpha$ line resemble well the 
         darkest absorption features present in the AIA $211$\,\AA\ image. The IRIS FOV,
         the area of
         possible SUMER slit occurrence within uncertainties, and the analyzed area of the 
         prominence -- that is the section along the solar\_\kern-0.7ptY, from which data were
         taken for the analysis -- are marked in all four images.}
\label{Fig:four_imgs_coal}
\end{figure*}

For comparison of the observed spectra with synthetic spectra produced by modeling
it is necessary to take into account that observed profiles are affected by 
instrumental broadening. To test how much the observed profiles are instrumentally 
broadened, we deconvolved instrumental profile from several selected \ion{Mg}{ii} 
k and h profiles of different shapes and widths observed in the prominence. For 
instrumental profile of IRIS NUV spectral band (both \ion{Mg}{ii} lines occur within 
this band), the Gaussian function of the Full Width at Half Maximum (FWHM) of 
$50.54\,\mathrm{m\AA}$ was used as in \citet{2020ApJ...888...42T}. For the 
deconvolution, the Wiener deconvolution numerical method (using the Fast Fourier 
Transform -- FFT -- with filtering of high frequencies in power spectra) was applied. 
After the deconvolution of selected profiles of different types and widths, 
the specific intensities changed remarkably unlike in the case of the SUMER 
Lyman\discretionary{-}{-}{-}line observations  (Sect.~\ref{Sec:sumer}). This 
is because the FWHM of the IRIS instrumental profile is larger by $77$\,\% than
that of SUMER. Moreover, the \ion{Mg}{ii} lines are narrower, on average by 
$2.4$ times, than profiles of the hydrogen Lyman lines Ly$\alpha$\,--\,Ly$\delta$. 
Because of the complexity and instability of the numerical methods for deconvolution 
of spectral profiles, we decided not to apply the deconvolution to the observed 
\ion{Mg}{ii} line profiles. Instead, the convolution with the instrumental profile, 
which is more numerically stable and mathematically simpler, will be applied to 
synthetic profiles produced by modeling in our next paper.
%
\subsection{Co-alignment of the SUMER and IRIS observations}\label{Sec:coalignment} 
Sophisticated co\discretionary{-}{-}{-}alignment of the SUMER slit and IRIS raster 
positions was carried out in several steps. First, shape and size of the prominence 
as seen in the IRIS \ion{Mg}{ii} $2796$\,\AA\ SJ image (hereafter referred to as IRIS
\ion{Mg}{ii} SJ image) was fitted to its appearance in the AIA $304$\,\AA\ image. 
This was made by matching of intensity contours plotted
to these two images for different X and Y mutual offsets and different angles of
their mutual rotation. It was found that contours match the best for no offsets
and no mutual rotation. Thus, this means that the IRIS SJ and AIA $304$\,\AA\ images
are well co\discretionary{-}{-}{-}aligned. It is because the coordinate system of
the IRIS instrument pointing is fixed according to the
full\discretionary{-}{-}{-}disk AIA observations (accuracy of the pointing stability
using the IRIS guide telescope is better than $1\,\mathrm{arcsec}$, see 
\citeauthor{2014ApJ...792L..15P}\ \citeyear{2014ApJ...792L..15P}). Thus, comparison
of the IRIS \ion{Mg}{ii} SJ and AIA $304$\,\AA\ images obtained at the times of the
nine analyzed IRIS raster observations showed that general shape and position of the
prominence is very similar in both of them except of its less bright and rather
dynamical peripheries, where shape of the prominence was changing remarkably within
time difference between the AIA and IRIS observations. Examples of the IRIS
\ion{Mg}{ii} SJ and AIA $304$\,\AA\ images obtained during beginning of the IRIS
prominence observations are shown in the upper row in Fig.~\ref{Fig:four_imgs_coal}.
In the second step, the H$\alpha$ filtrogram observations of the prominence obtained
with the High Resolution H$\alpha$ Imaging System
\citep{2003HvaOB..27..189O} at the Kanzelh\"{o}he Solar Observatory (KSO)
were co\discretionary{-}{-}{-}aligned with the AIA image in the 211\,\AA\ 
channel. The dark structure visible in the AIA 211\,\AA\ image shown in lower left panel of 
Fig.~\ref{Fig:four_imgs_coal}, is an absorption counterpart of the prominence. It is 
caused by absorption of the EUV coronal radiation in re\-so\-nan\-ce continua of 
hydrogen and helium and coronal emissivity deficit (see, e.g., 
\citeauthor{2005ApJ...622..714A} \citeyear{2005ApJ...622..714A} 
and \citeauthor{2015A&A...574A..62S} \citeyear{2015A&A...574A..62S}). It is clearly 
distinguishable that its general shape resembles well the prominence in emission as 
seen in the H$\alpha$ filtergram (lower right panel of Fig.~\ref{Fig:four_imgs_coal}).  
It was not necessary to perform any special co\discretionary{-}{-}{-}alignment of 
the AIA $211$\,\AA\ and AIA $304$\,\AA\ images because AIA observations from different
channels are co\discretionary{-}{-}{-}aligned during data reduction by the Solar Soft procedure 
\texttt{aia\_prep.pro}. Thus, after completing the second step, the H$\alpha$ KSO filtergram 
(Fig.~\ref{Fig:four_imgs_coal} lower right panel) is co\discretionary{-}{-}{-}aligned with 
the IRIS \ion{Mg}{ii} SJ image (Fig.~\ref{Fig:four_imgs_coal} upper left panel) and 
subsequently with the \ion{Mg}{ii} line rasters as position and FOV of the raster can be 
fixed exactly by tracking position of the spectrograph slit visible as dark vertical 
line in SJ images. The IRIS FOV is marked by the blue dashed line in all four 
panels of Fig.~\ref{Fig:four_imgs_coal}. 

In the final step of the co\discretionary{-}{-}{-}alignment, position of the SUMER 
spectrograph slit (hereafter referred simply to as slit position) in the IRIS \ion{Mg}{ii} 
SJ image was estimated. This task is rather complicated because SUMER  
records no slit\discretionary{-}{-}{-}jaw images, which could be easily fitted to the 
filtergrams shown in Fig.~\ref{Fig:four_imgs_coal}. Thus, the only possibility was 
searching for correct slit position by 
trial\discretionary{-}{-}{-}and\discretionary{-}{-}{-}error approach placing 
it in different locations (without changing its length and roll angle) within an area of 
the whole FOV of IRIS \ion{Mg}{ii} SJ image and evaluating cross\discretionary{-}{-}{-}correlation 
(hereafter referred to as KSO\discretionary{-}{-}{-}SUMER correlation coefficient) between 
distribution of H$\alpha$ intensities from the KSO filtergram along the slit position and
integrated intensities of the Ly$\alpha$ observed by SUMER. 
Only positions along the spectrograph
slit between $30$ and $59\,\mathrm{arcsec}$ (measured from the solar south)
were used in calculations of the correlation coefficient
in order to avoid locations with a lot of dynamics where data are varying remarkably with
time --  locations close to the solar limb where very dynamical spicules occurred.
Either positions at low\discretionary{-}{-}{-}dense and very dynamical peripheries
of the prominence (above $59\,\mathrm{arcsec}$) are
not suitable for including into
the co\discretionary{-}{-}{-}alignment. The section along the slit taken for the
co\discretionary{-}{-}{-}alignment is bordered by two blue horizontal lines both in
the intensity maps and example spectra in Fig.~\ref{Fig:SUMERobs}. Vertical positions
in this figure are given in the
helioprojective\discretionary{-}{-}{-}Cartesian Y coordinates (solar\_\kern-0.7ptY)
for the SUMER slit location determined by the co\discretionary{-}{-}{-}alignment.
Solely data from section between these positions is taken into the analysis. This is
described in more detail in the next paragraph. The highest
KSO\discretionary{-}{-}{-}SUMER correlation coefficient of $0.89$ was obtained
for the slit position located at 
\mbox{$\mathrm{solar\_X}=817.0\,\mathrm{arcsec}$} and
\mbox{$\mathrm{solar\_}\kern-0.7pt\mathrm{Y}=608.0\,\mathrm{arcsec}$} (slit center).
An effort for exact co\discretionary{-}{-}{-}alignment based on 
cross\discretionary{-}{-}{-}correlation of 1D data (the Ly$\alpha$ integrated intensities 
along the SUMER slit) within 2D data (KSO H$\alpha$ filtergram) cannot be reliable, although
rather high value of the KSO\discretionary{-}{-}{-}SUMER correlation coefficient was
obtained for the above mentioned SUMER slit position. Therefore, it is much more correct in such case to use the co\discretionary{-}{-}{-}alignment to estimate
of an area which encompasses
possible slit positions with acceptable values of the
KSO\discretionary{-}{-}{-}SUMER correlation coefficient instead of trying to find an
exact slit position. The spectrograph slit position was being gradually shifted in both
solar\_X and solar\_\kern-0.7ptY directions on both sides from the position with the highest value
of the KSO\discretionary{-}{-}{-}SUMER correlation
coefficient. The correlation coefficient was decreasing slowly at first and after reaching 
the value of $\sim\,0.8$ it suddenly dropped down to much lower values. Because a value of 
the correlation coefficient of $0.8$ is sill acceptable, an area of possible slit positions 
was fixed encompassing those with the KSO\discretionary{-}{-}{-}SUMER correlation coefficient 
equal or higher than $0.8$. The area is marked by green line in all four panels of 
Fig.~\ref{Fig:four_imgs_coal}. The area overlaps at least partially with the IRIS raster. 
Thus, the SUMER slit might lie within the IRIS raster, but it is also possible that it 
occurred outside but not far. Because the studied prominence appears rather homogeneous 
within the region of interest (marked by two orange horizontal lines in 
Fig.~\ref{Fig:four_imgs_coal}), such large co\discretionary{-}{-}{-}alignment uncertainty 
should not play a significant role in our analysis. 

As was already said in the previous paragraph, only data from the section between
positions $30$ and $59\,\mathrm{arcsec}$  (measured from the solar south) along
the SUMER slit was used for the co\discretionary{-}{-}{-}alignment. This section
corresponds
to the solar\_\kern-0.7ptY positions between $575$ and $604$\,arcsec for the slit
position with the highest KSO\discretionary{-}{-}{-}SUMER correlation coefficient.
The observations of SUMER and IRIS do not overlap in time, therefore we decided to take
only this section along the solar\_\kern-0.7ptY also into our analysis. Hereafter, this
section along the solar\_\kern-0.7ptY is called analyzed area of the prominence. One can
clearly see in the maps of integrated intensities constructed for the observations of
SUMER and IRIS \,--\,Figs.~\ref{Fig:SUMERobs} and \ref{Fig:IRISobs}\,--\,that intensity
structures inside of this area (bordered by the two blue horizontal lines in both figures)
are relatively stable in time, while there is really a lot of dynamics outside of it.
%
%
\section{Methods of the statistical spectroscopic analysis}\label{Sec:method} 
\subsection{Classification of profiles according to number of peaks}\label{Sec:profpeaks} 
Emission profiles of optically thick spectral lines emitted from the cool prominence 
plasma and formed by the non\discretionary{-}{-}{-}LTE scattering are often reversed 
in their cores. Moreover, in the case of gradients of LOS velocities of the prominence
plasma along LOS, absorption profiles are
Doppler\discretionary{-}{-}{-}shifted accordingly which can produce profiles with
multiple peaks and reversals (see, e.g.,
\citeauthor{2020ApJ...888...42T}\ \citeyear{2020ApJ...888...42T}). 
The method for identification of peaks in profiles used in this work is based on  
searching for local maxima in the profile core. Firstly, profile center  
is determined by fitting its wings with the Voigt function. Then, the profile is 
searched for local maxima from the center toward both wings until $30$\,\% of
the value of global maximum in the profile is reached. To eliminate the influence of 
noise, profiles were smoothed by running average before searching for peaks. 
The running average was performed in such a way that for each wavelength point 
specific intensity in this point is taken with weight of unity and intensities 
of adjacent points are added with weights of $0.5$. Even after the averaging, 
only peaks with intensities higher than $5$\,\% above the adjacent dips are 
considered. 

There is also a possibility of evaluation of credibility of each detected peak -- 
whether it is ``\,above error bars\,'' -- which we apply in Sect.~\ref{Sec:results}.
The condition for this
is that the specific intensity after subtraction of measurement error should be higher
than intensities at adjacent reversals after adding measurement errors to them. If it
is not the case, the peak is assumed as ``\,lost in noise\,''. Unlike the peak identification,
profiles with original (not smoothed) measured specific intensities with errors at all
wavelength points and no other conditions except that above described, are used for
evaluation of peak credibility.
%
%
\subsection{Profile characteristics}
\label{Sec:profchars}
\begin{table*}
\caption{Designation of quantities used in statistics of the profile characteristics.}
\label{tab:statquantities}
\centering 
\begin{tabular}{llc} 
\hline
\hline
quantity\hspace{30ex}   &           detailed description\hspace{30ex}    &    designation \\
\hline
                        & wavelength integrated intensities after       &                 \\[-1.4ex]
integrated intensity    &                                               &            $E$  \\[-1.4ex]
                        &  \hspace{1ex}removing of influences of blends &  \\[0.9ex]
blue\discretionary{-}{-}{-}peak intensity & specific intensity at the profile blue peak &    
  $I_{\mathrm{b}}$ \\[0.9ex]
                          & intensity in a dip occurring in the profile      &       \\[-1.4ex]
intensity in the reversal &                      &   $I_{\mathrm{rev}}$ \\[-1.4ex]
                        &  due to self\discretionary{-}{-}{-}absorption &        \\[0.9ex]
red\discretionary{-}{-}{-}peak intensity & specific intensity at the profile red peak & 
  $I_{\mathrm{r}}$ \\[0.9ex]
depth of the central reversal  & ratio of intensity $I_{\mathrm{rev}}$ to average of &                  
                    \\[-1.4ex]
                              &                                          &  $r_{\mathrm{CP}}$ \\[-1.4ex] 
(also center-to-peaks ratio)  & the $I_{\mathrm{b}}$ and $I_{\mathrm{r}}$ intensities &  
          \\[0.9ex]
peak asymmetry & ratio of the peak intensities $I_{\mathrm{b}}$ and $I_{\mathrm{r}}$ & $r_{\mathrm{PA}}$ \\[0.9ex]
blue\discretionary{-}{-}{-}to\discretionary{-}{-}{-}RED peak ratio & $I_{\mathrm{b}}/I_{\mathrm{r}}$  
 when $I_{\mathrm{b}}<I_{\mathrm{r}}$ & $r_{\mathrm{B-R}}$ \\[0.9ex]
red\discretionary{-}{-}{-}to\discretionary{-}{-}{-}BLUE peak ratio & $I_{\mathrm{r}}/I_{\mathrm{b}}$  
 when $I_{\mathrm{r}}<I_{\mathrm{b}}$ & $r_{\mathrm{R-B}}$ \\[0.9ex]
\hline 
\end{tabular}
\end{table*}
In our spectroscopic analysis, we used the same profile characteristics 
as in \citet{2010A&A...514A..43G} and \citet{2015A&A...577A..92S} except for the Lyman 
decrement. The Lyman decrement is defined as the ratio of integrated intensities of a 
given line to that of Ly$\beta$. 
However, unlike in \citet{2010A&A...514A..43G} and \citet{2015A&A...577A..92S}, the 
Lyman\discretionary{-}{-}{-}line data used here was obtained in blocks and not
sequentially (see Sect.~\ref{Sec:sumer} for more details). Time delays between 
observations of individual spectral lines are thus much larger -- around $10$ minutes 
instead of several seconds. We therefore decided not to use the Lyman 
decrement for the spectral analysis. Designations of all quantities used for calculations 
of profile characteristics together with their detailed description, are in 
Table~\ref{tab:statquantities}.

The first profile characteristics is the integrated intensity $E$. Wavelength 
interval of integration is determined in such a way that whole emission part 
of the spectral line profile is integrated and in this way blends in the far wings 
are excluded. This profile characteristics can be used for all types of profiles. 

The second characteristics is the depth of central reversal $r_{\mathrm{CP}}$ (called  
center\discretionary{-}{-}{-}to\discretionary{-}{-}{-}peaks ratio
in our previous works). 
To minimize the influence of noise, average values from three adjacent 
wavelength points in the reversal and at each peak were considered. This profile 
characteristics is defined only for 2\discretionary{-}{-}{-}peak profiles.

The third profile characteristics used in the analysis is the peak asymmetry $r_{\mathrm{PA}}$. 
This is also defined only for 2\discretionary{-}{-}{-}peak profiles. For better intelligibility 
of statistics of the peak asymmetry, profiles with higher red peak and with higher blue peak are 
analyzed separately (see $r_{\mathrm{B-R}}$ and $r_{\mathrm{R-B}}$ in Table~\ref{tab:statquantities}). 
Symmetric profiles are also treated individually because 
of their much higher count than asymmetric profiles. Influence of noise is again minimized by 
averaging over three adjacent spectral points in each peak. 

As was shown in our previous works \citep{2010A&A...514A..43G, 2015A&A...577A..92S},  
these profiles characteristics are suitable for statistical comparison of observed 
Lyman\discretionary{-}{-}{-}line profiles with synthetic ones obtained by 
non\discretionary{-}{-}{-}LTE modeling. The same three profile characteristics are
also used for the \ion{Mg}{ii} k and h lines. 

We note that other, more advanced methods are also suitable for the classifications
of line profiles, such as Principal Component Analysis
(PCA; \citeauthor{cit:pcamethod} \citeyear{cit:pcamethod}) and
k\discretionary{-}{-}{-}mean clustering \citep{cit:kmeanclusteringmethod}. For example,
PCA was used by \citet{2020AN....341...64D} for
cloud\discretionary{-}{-}{-}model diagnostics of a prominence plasma from observed
H$\alpha$ profiles, k\discretionary{-}{-}{-}mean clustering for analysis of Doppler
velocities in a filament from profiles of H$\alpha$ and \ion{Ca}{ii}~$8542$\,\AA\ by
\citet{2015ApJ...806....9K}. However, in the current paper we apply the methods used
in our previous works to maintain continuity and inter comparability.
%
\subsection{Influence of blends} 
\label{Sec:blends_infl}
\begin{table}
\caption{Contributions of blends in percents (CBP) to quantities used in calculations
of the profile characteristics for the six spectral lines.}
\label{tab:BlendContribProfChars}
\centering 
\begin{tabular}{lcccc}
\hline
\hline\\[-2.0ex]
sp. line & CBP to $E$ & CBP to $I_{\mathrm{b}}$ & CBP to $I_{\mathrm{rev}}$ & CBP to $I_{\mathrm{r}}$ \\
\hline
                                &                                             &      &      &     \\[-2.0ex]
Ly$\alpha$                      & \hspace{1ex}$0$                           & $0$  & $0$  & $0$ \\
Ly$\beta^{\hspace{0.5ex}\ast}$  & \hspace{1ex}$4$                           & $0$  & $1$  & $1$ \\
Ly$\gamma$                      & \hspace{1ex}$3$                           & $0$  & $0$  & $1$ \\
Ly$\delta$                      & $37$                                      & $2$ & $0$ & $2$ \\
\ion{Mg}{ii} k                  & \hspace{2.0ex}$0^{\hspace{0.2ex}\dagger}$ &  $0$ &  $0$ & $0$  \\
\ion{Mg}{ii} h                  & \hspace{1ex}$0$                           &  $0$ &  $0$ & $0$  \\[0.2ex]
\hline 
\end{tabular}
\tablefoot{Quantities in individual columns: CBP to integrated intensities \\
           \phantom{xxxxxx}\hspace{0.8ex}in the second
           column, CBP's to specific intensities at the blue\\
           \phantom{xxxxxx}\hspace{0.8ex}peak, central reversal, and red peak
           in further three columns.\\[0.5ex]
           \phantom{xxxxx}\hspace{0.7ex}$^{\ast}$ influence of \ion{O}{i} $1025.76$\,\AA\ is not taken into account\\
           \phantom{xxxxx}\hspace{0.6ex}$^{\dagger}$ \ion{Mn}{i} $2795.6$\,\AA\ is outside the
           wavelength interval of the integra\\
           \phantom{xxxxx}\hspace{2.5ex}-tion} 
\end{table}
\begin{figure*}
\centering
\resizebox{\hsize}{!}{
\includegraphics{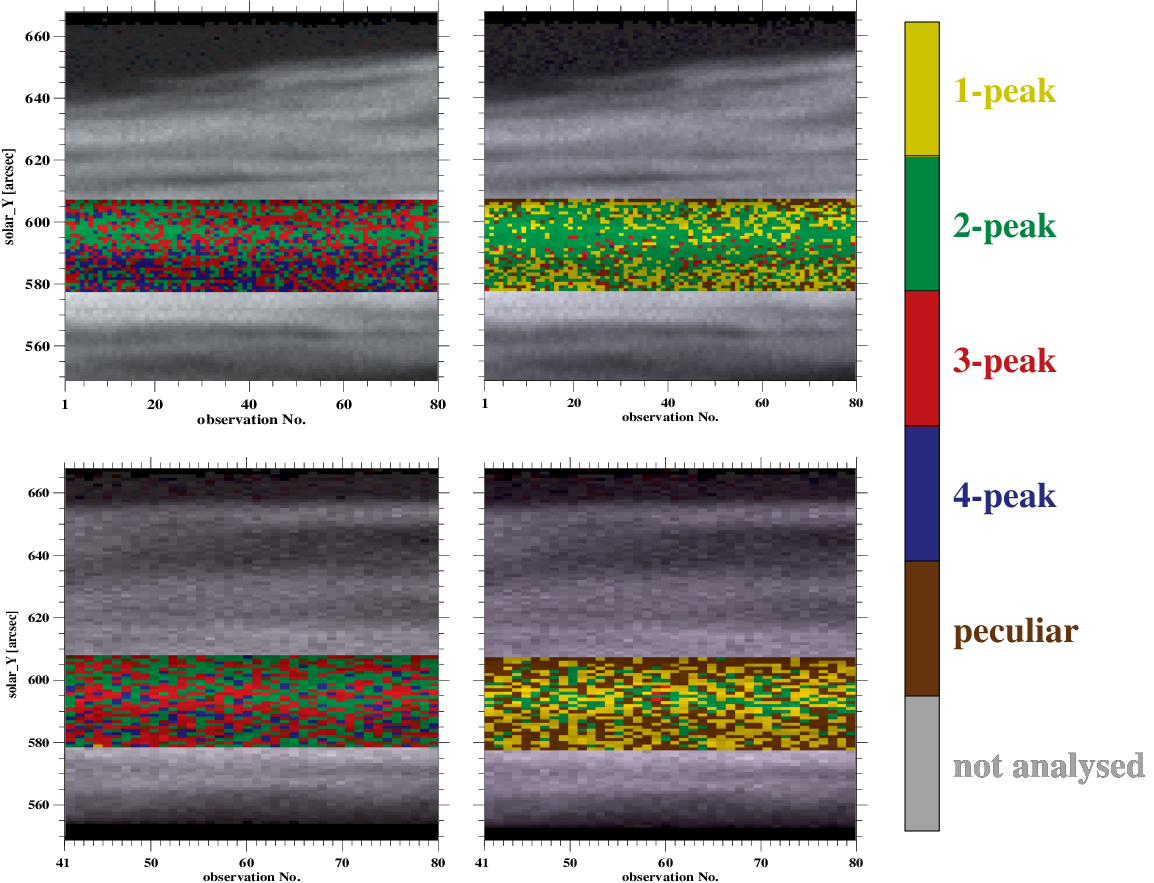}} 
\caption{Maps of the integrated intensities of Ly$\alpha$ (top row) and Ly$\beta$ (bottom 
row) overlaid with the colored mosaics classifying profile types (color bar on the right) 
within the target area. Profile types are classified regardless of noise (left column) 
and by considering only the peaks safely above error bars (right column). All available Ly$\alpha$
data were employed (see Fig.~\ref{Fig:SUMERobs} panel b) while only the Ly$\beta$ data with
the final pointing were considered in the mosaics (Table~\ref{tab:SumObs}).
} 
\label{Fig:res_proftypemaps}
\end{figure*}
\begin{figure*}
\centering
\resizebox{\hsize}{!}{
\includegraphics{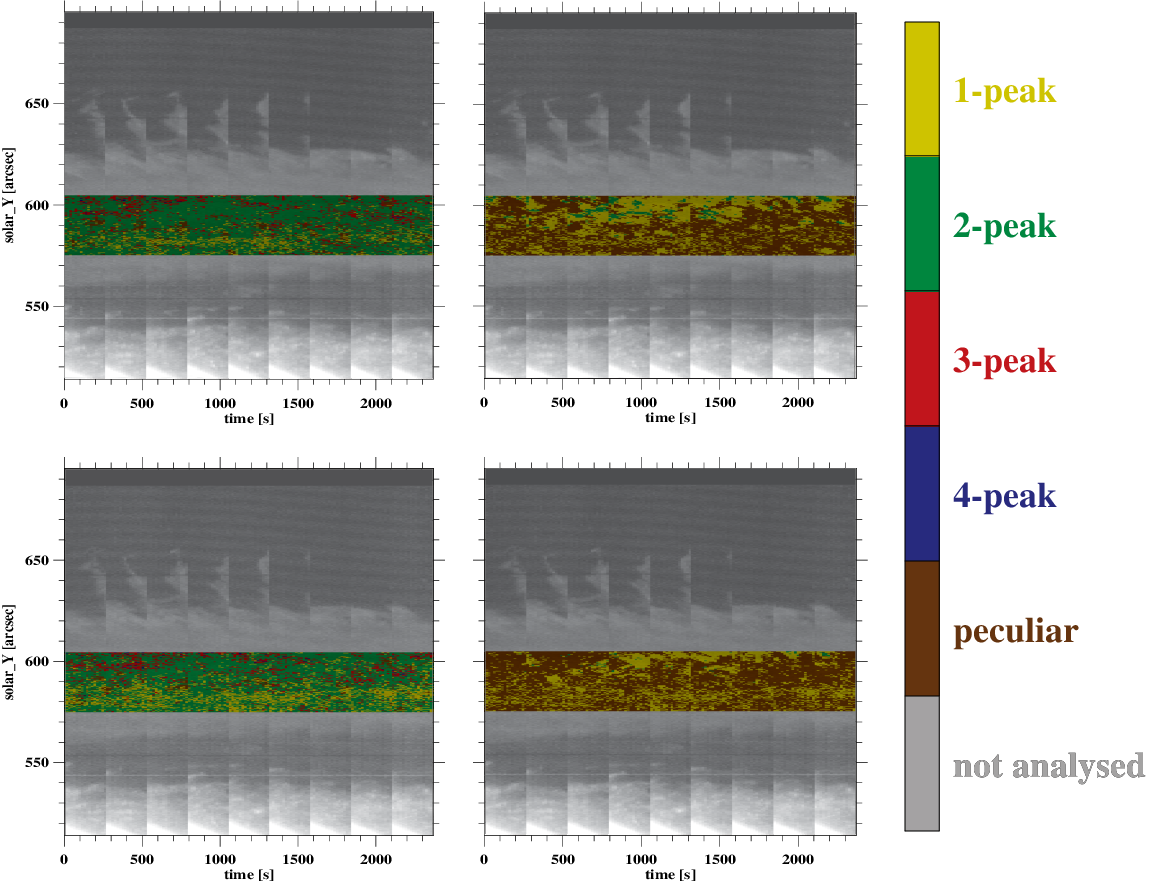}
} 
\caption{
Maps of the integrated intensities of the \ion{Mg}{ii} k (top row) and \ion{Mg}{ii} h lines  
(bottom row) overlaid with the colored mosaics classifying profile types (color bar 
on the right) within the target area. Here only the profile types from first nine 
rasters are classified regardless of noise (left column) and by considering only 
the peaks above error bars (right column).
} 
\label{Fig:res_proftypemaps1}
\end{figure*}
\begin{figure*}[p]
\centering
\resizebox{0.88\hsize}{!}{
\includegraphics{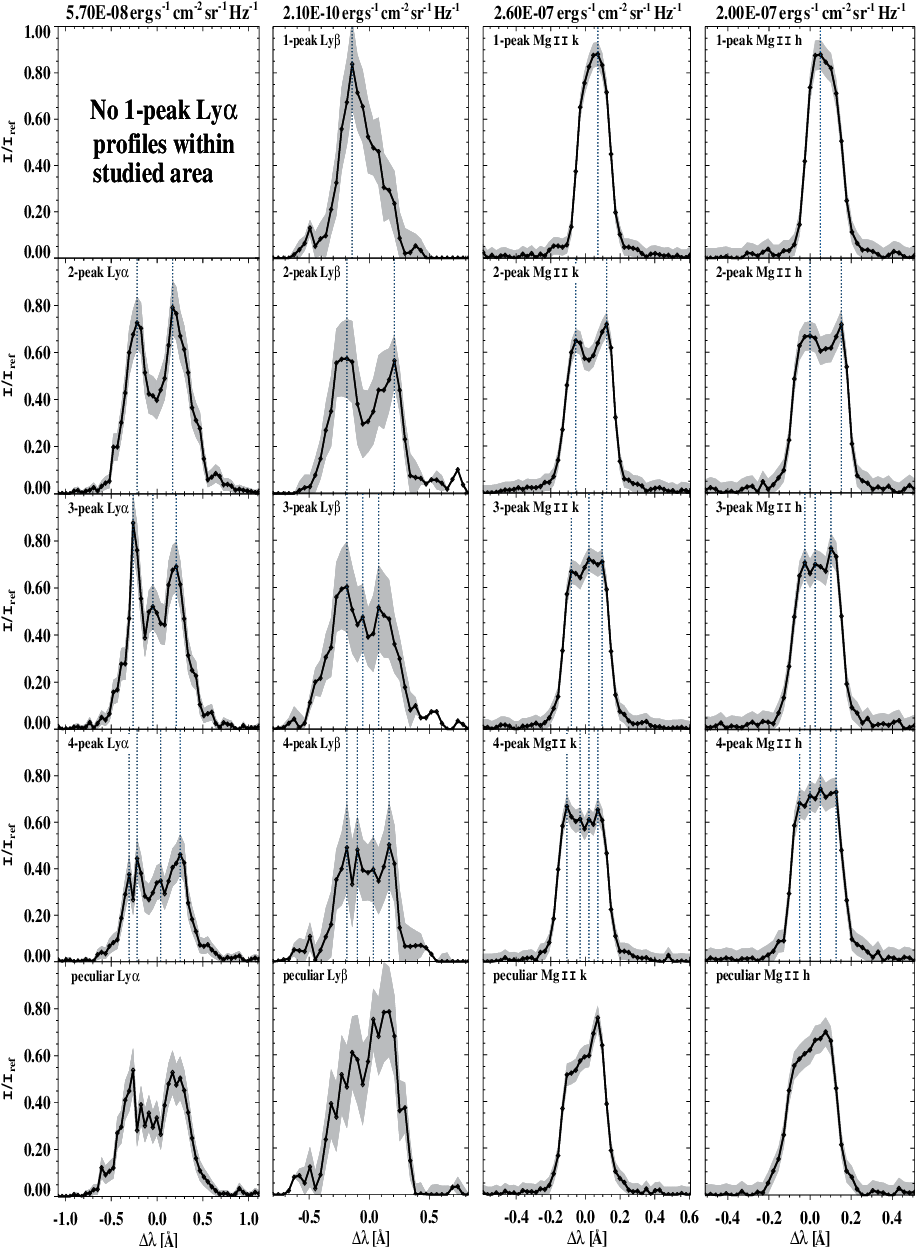}}
\caption{Examples of different types of spectral line profiles according to the number of 
peaks detected regardless of the noise (as in colored maps shown in the columns on the left
in Figs.~\ref{Fig:res_proftypemaps} and \ref{Fig:res_proftypemaps1}).
In columns from the left, spectral profiles of Ly$\alpha$, Ly$\beta$, \ion{Mg}{ii} k,
and \ion{Mg}{ii} h are shown. In rows from the top, plots of 1-, 2-, 3-,
4\discretionary{-}{-}{-}peak and peculiar profiles are located. Error margins of intensities
are shown as gray areas around plot\discretionary{-}{-}{-}lines. On abscissas, relative
wavelength scales $\Delta\lambda=\lambda-\lambda_{\mathrm{0}}$ are used (for the
$\lambda_{\mathrm{0}}$ values see Table~\ref{tab:LineList}). The relative intensities 
$I/I_{\mathrm{ref}}$ are on ordinates; intensities 
$I_{\mathrm{ref}}$ in 
$\mathrm{erg}\,\mathrm{s}^{-1}\mathrm{cm}^{-2}\mathrm{sr}^{-1}\,\mathrm{Hz}^{-1}$ 
are shown at the top of each column for individual spectral lines. The positions of all 
peaks  are marked with blue vertical 
dotted lines. For a given type of profile (row of the figure) the Lyman line profiles are 
not from the same position nor from the same time while \ion{Mg}{ii} k and h profiles 
of given type are.}  
\label{fig:profexmpls}
\end{figure*}
%
\begin{figure*}[h!]
\parbox{0.90cm}{\phantom{xx}}\
\parbox{0.75\hsize}{
\resizebox{\hsize}{!}{
\includegraphics{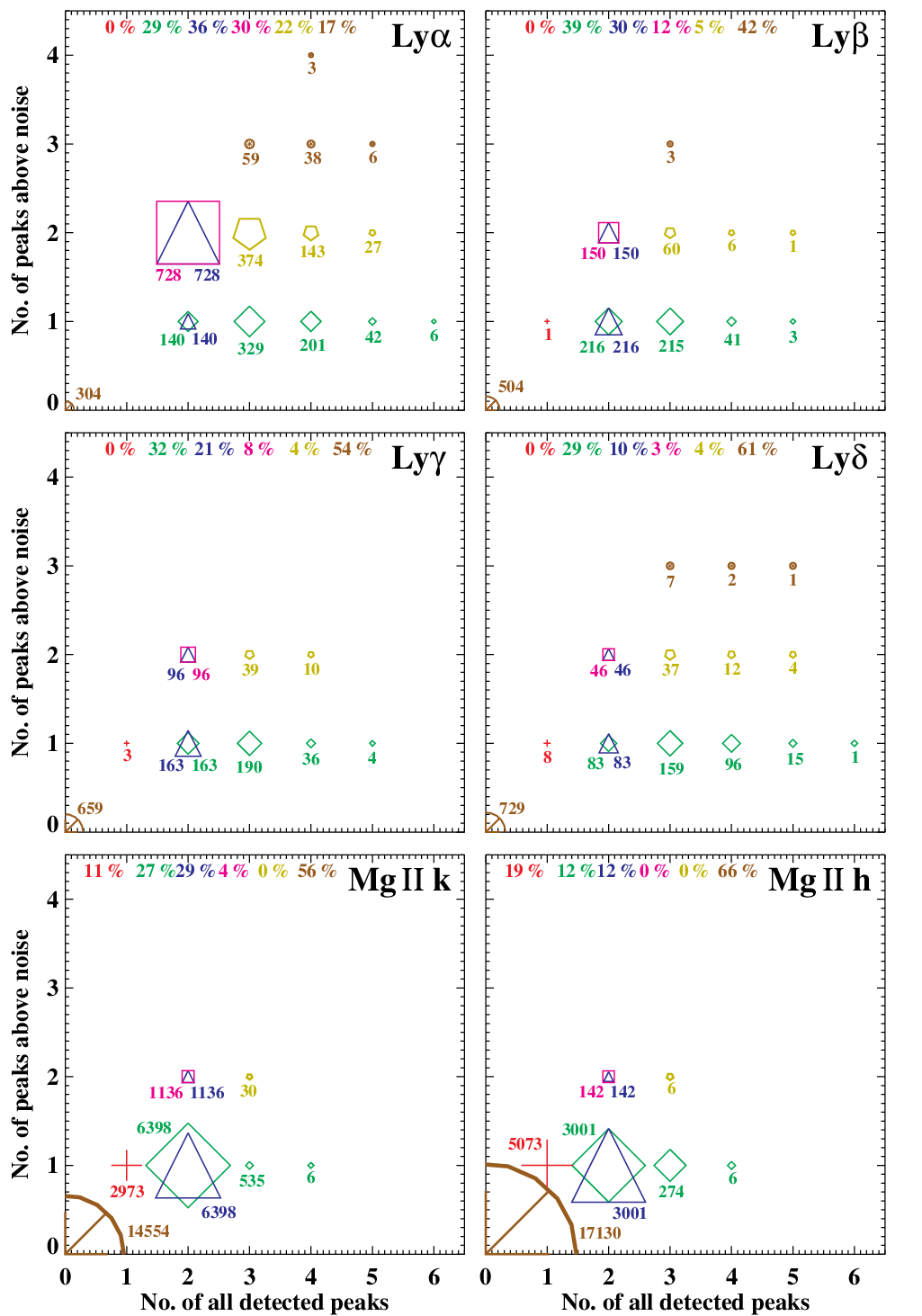}}}
\caption{Diagrams showing distributions of profiles of all six spectral lines in a space  
of the number of all of the detected peaks on the abscissa versus number of peaks above
noise on the ordinate.
Affiliations to the individual groups are indicated by different colors and symbols; the 
legend is located on the right. Sizes of the symbols are proportional 
to numbers of profiles in the individual positions and these numbers are also shown next 
to the symbols. Further information on the diagrams is given in the text. 
}
\label{Fig:res_profgroupsplots}
\end{figure*}
%
For precise determination of the profile characteristics, the influence of blends needs
to be taken into account. We thus removed their contribution in all profiles included
into the analysis. Contributions of blends were estimated similarly as in
\citet{2015A&A...577A..92S}. Profiles of all six lines averaged from analyzed area of
the prominence were fitted by the sum of the Voigt profile and negative Gaussian to
reproduce two peaks and reversal in between occurring in the average profiles. Blends
are then fitted simply with Gaussians and their contributions are then calculated using
these Gaussian fits. The averaged profiles were used to evaluate percentual
contributions of blends, although evaluation of influence of blends would be also
possible by fitting profiles from each position and time individually. But due to
occurrence of multi\discretionary{-}{-}{-}peak features in many of them caused by
large dynamics and/or noise,
fitting of blends could be problematic and contributions to the
individual profile characteristics resulting from the fitting could by biased. Average profiles
have simple shapes with clearly pronounced two peaks and
reversal without any visible manifestation of noise and therefore are more suitable for
a reliable fitting. For integrated intensities $E$, contribution of blends is calculated
by summation of integrals of their Gaussian fits. Contributions of the blends to specific
intensities at the blue peak, central reversal, and red peak in analyzed profiles are
taken from values of their Gaussian fits at these three wavelength points. The
contributions of blends in percents to integrated intensities and to specific intensities
at the three wavelength points for observed spectral lines are listed in
Table~\ref{tab:BlendContribProfChars}.

For Ly$\alpha$, no blends were identified within observed wavelength range. 
In the case of Ly$\beta$, three blends -- \ion{He}{ii} $1025.27$\,\AA,
\ion{O}{i} $1025.76$\,\AA, and \ion{Fe}{iii} $1026.76$\,\AA\ -- are present according to the SUMER
spectral atlas \citep{2001A&A...375..591C}. The \ion{He}{ii} and \ion{Fe}{iii} lines were well  
distinguishable in the average profile and fitted and thus corrections to the profile 
characteristics due to these blends were successfully estimated and applied.  
The \ion{O}{i} $1025.76$\,\AA\ line is too close to the Ly$\beta$ core to be reliably
fitted and therefore its contributions to the profile characteristics could 
not be evaluated. But it cannot be neglected, because according to the work of
\citet{1997A&AS..126..281C}, upper level of transition emitting this line is 
in part photo\discretionary{-}{-}{-}pumped by the Ly$\beta$ photons. This can lead to 
remarkable increase of its brightness. 
Although this effect was studied in detail by \citet{1993ApJ...402..344C}, unfortunately  
they have not evaluated specific intensities of the \ion{O}{i} $1025.76$\,\AA\ line
relatively to Ly$\beta$. Therefore, contribution of this blend could not be determined
and no corrections due to this line were applied. But its presence within the Ly$\beta$
profiles should be kept in mind when results of our analysis are compared with the
non\discretionary{-}{-}{-}LTE modeling.
For the Ly$\gamma$, profile characteristics were corrected for the blending lines 
\ion{He}{ii} $972.12$\,\AA\ and \ion{O}{i} $973.24$\,\AA\ occurring in its blue and red wings,
respectively. In the case of the Ly$\delta$, there is a strong influence of the blends 
\ion{He}{ii} $949.36$\,\AA, \ion{Si}{viii} $949.22$\,\AA, and
\ion{Fe}{iii} $950.34$\,\AA\ on the line profile as can be seen in the
Table~\ref{tab:BlendContribProfChars}.
Other blends of \ion{O}{i} and \ion{Si}{ix} occur within the Ly$\delta$ line core, 
thus it was not possible to fit them reliably. The \ion{O}{i} lines are 
photo\discretionary{-}{-}{-}pumped by the Ly$\delta$ photons \citep{1997A&AS..126..281C} 
similarly, as it was for the \ion{O}{i} line blending Ly$\beta$.
Unfortunately, contributions of these possibly prominent blending lines to
Ly$\delta$ cannot be spe\-ci\-fied, because estimations of their intensities lack.
But  nevertheless, occurrence of
these blends within the Ly$\delta$ profiles should be kept in mind when comparing with
the modeling. According to the SUMER spectral atlas \citep{2001A&A...375..591C},
the \ion{Si}{ix} lines occurring within the Ly$\delta$ profile are faint and their
contributions to the Ly$\delta$ profile characteristics can be neglected.

According to \citet{2015ApJ...811..127S}, 
there is a notable blend \ion{Mn}{i} $2795.6$\,\AA\ in the blue wing of the \ion{Mg}{ii}
k line. Fortunately, \ion{Mg}{ii} k line profiles are narrow enough that the blend 
occurs in the wing far from the line. We could thus neglect it. In the case of the 
\ion{Mg}{ii} h line, no blends were identified within the wavelength interval of 
its emission profile. 
%
\section{Results}\label{Sec:results} 
\subsection{Statistics of profile classification}\label{Sec:results_proftypes} 
Peaks in profiles of all six lines within the analyzed area of the prominence were
detected regardless noise by the method described in Sect.~\ref{Sec:profpeaks}.
Profiles were classified as 1\discretionary{-}{-}{-}peak profiles
(also referred to as purely emission profiles), and 2\discretionary{-}{-}{-}peak,
3\discretionary{-}{-}{-}peak, and 4\discretionary{-}{-}{-}peak profiles.
Those with unusual shapes without clearly distinguishable peaks by the method are
con\-si\-de\-red as peculiar.
Peculiar profiles also include profiles shaped as though
they had their top cut off, often containing one narrow unrealistically sharp
peak located remarkably asymmetrically from their centers.
There are usually also sort of hints of more possible peaks.
Although such profiles could be assumed as 1\discretionary{-}{-}{-}peak,
the method can distinguish them as peculiar.
There is most likely more than one peak
but those peaks (manifesting themselves just as hints) are too small to rise clearly
above adjacent reversals and thus, they cannot be detected.
Profiles with more than four
distinguishable peaks are not realistic due to large optical thickness in cores of
the six lines. For typical LOS velocities up to $20$\,km\,s$^{-1}$
occurring in quiescent prominences, such profiles cannot arise. Moreover, their
statistical occurrences within the analyzed area are less than $1$\,\%,
and their locations are rather scattered (they are not localized in any compact
areas). Therefore, these profiles were also included among the peculiar ones.
Maps of the integrated intensities were constructed with types of profiles
according to number of peaks indicated by different colors in each pixel
within the analyzed area.
Positions in the prominence outside of the analyzed area (see the last paragraph of  Sect.~\ref{Sec:coalignment}) are indicated by gray color. The maps for Ly$\alpha$ and 
Ly$\beta$ are shown in the left column of Fig.~\ref{Fig:res_proftypemaps} and for
the \ion{Mg}{ii} k and h lines in the left column of Fig.~\ref{Fig:res_proftypemaps1}. 
These maps are hereafter referred to as all\discretionary{-}{-}{-}peak maps.
In the maps shown in right columns of Figs.~\ref{Fig:res_proftypemaps} and
\ref{Fig:res_proftypemaps1}, classification of profiles was done taking
into account only peaks above error bars.
These maps are hereafter called as above\discretionary{-}{-}{-}error maps.

When inspecting the Ly$\alpha$ all\discretionary{-}{-}{-}peak map (upper left panel 
of Fig.~\ref{Fig:res_proftypemaps}), one can see that there are all types of profiles
except 1\discretionary{-}{-}{-}peak profiles. No clear trend in distributions of different 
profile types is evident. It can be just stated that the 2\discretionary{-}{-}{-}peak 
profiles occur mainly above solar\_Y of  approx. $595\,\mathrm{arcsec}$ forming 
several small areas. Below this solar\_Y position, the map is dominated by 3- and 
4\discretionary{-}{-}{-}peak profiles. In the Ly$\alpha$ 
above\discretionary{-}{-}{-}error map (right upper panel of the figure), there are 
almost only 1\discretionary{-}{-}{-}peak and 2\discretionary{-}{-}{-}peak profiles with
clearly visible trend of their occurrence. The large number of 1\discretionary{-}{-}{-}peak
Ly$\alpha$ profiles in such a strong line with
characteristically pronounced self\discretionary{-}{-}{-}absorption is caused 
by low intensity of one of the peaks, which thus becomes lost in noise. This fact
emphasizes the complex issue of the influence of noise on the statistical analysis of 
even such intense line as Ly$\alpha$. The 2\discretionary{-}{-}{-}peak profiles are 
gathered mainly in the central part of the analyzed area (section along the slit 
between approx. $\mathrm{solar}\_\kern-0.7pt\mathrm{Y}=590$ and $600\,\mathrm{arcsec}$).
At the edges of the analyzed area, the 1\discretionary{-}{-}{-}peak profiles dominate.
Proportion of peculiar profile is little bit higher than in the
all\discretionary{-}{-}{-}peak map.

In the case of Ly$\beta$, one can see 
in the all\discretionary{-}{-}{-}peak map (left panel in the second row of 
Fig.~\ref{Fig:res_proftypemaps}) that relative count of the 2\discretionary{-}{-}{-}peak 
profiles is smaller than for Ly$\alpha$ while higher portion of the 
3\discretionary{-}{-}{-}peak profiles is clearly visible. Neither here any clear trend 
is seen in distributions of different profile types. There is a small amount of the 
1\discretionary{-}{-}{-}peak profiles unlike in Ly$\alpha$ where no 
1\discretionary{-}{-}{-}peak profiles occur in the all\discretionary{-}{-}{-}peak map. 
Similarly as for Ly$\alpha$, there are only a few peculiar Ly$\beta$ profiles. 
In the Ly$\beta$ above\discretionary{-}{-}{-}error map (right 
panel in the second row of Fig.~\ref{Fig:res_proftypemaps}), 1\discretionary{-}{-}{-}peak 
and peculiar profiles clearly dominate. There are only few 2\discretionary{-}{-}{-}peak 
and no 3- and  4\discretionary{-}{-}{-}peak profiles. Similar all\discretionary{-}{-}{-}peak 
and above\discretionary{-}{-}{-}error maps as in the case of Ly$\beta$, were obtained 
also for Ly$\gamma$ and Ly$\delta$ (not shown in Fig.~\ref{Fig:res_proftypemaps}). 

There is majority of the 2\discretionary{-}{-}{-}peak \ion{Mg}{ii} k line profiles as can 
be seen in its all\discretionary{-}{-}{-}peak map in the left panel of the first row of 
Fig.~\ref{Fig:res_proftypemaps1}. There are several smaller areas with occurrence of purely 
emission profiles mainly up to $\mathrm{solar}\_\mathrm{Y}\approx590\,\mathrm{arcsec}$.
Count of the profiles
with more than 2\discretionary{-}{-}{-}peaks is rather low. But there are almost only purely 
emission and peculiar profiles in the above\discretionary{-}{-}{-}error map of this line. 
Purely emission profiles are gathered mainly in the area between times of approximately 
$500$\,--\,$1850\,\mathrm{s}$ and within positions along the slit above solar\_Y  of 
$590\,\mathrm{arcsec}$. Within this area, several small regions with occurrence of the 
2\discretionary{-}{-}{-}peak profiles are also located. The maps for the \ion{Mg}{ii} h 
displayed in the bottom row of Fig.~\ref{Fig:res_proftypemaps1} look similar, with 
purely emission profiles occurring also mainly up to 
$\mathrm{solar}\_\mathrm{Y}\approx590\,\mathrm{arcsec}$ in the all\discretionary{-}{-}{-}peak
map but in larger count than for the k line. In the above\discretionary{-}{-}{-}error map, 
there is larger count of peculiar profiles on expense of purely emission profiles and 
almost no 2\discretionary{-}{-}{-}peak profiles contrary to the k line 
above\discretionary{-}{-}{-}error map. Thus, comparing the all\discretionary{-}{-}{-}peak and 
above\discretionary{-}{-}{-}error maps for the both \ion{Mg}{ii} k and h lines indicates that 
majority of peaks in profiles within the analyzed area are not above error bars.

Examples of all types of profiles (including those assumed as peculiar profiles)
of the hydrogen Lyman and \ion{Mg}{ii} k and h  spectral lines are shown in
Fig.~\ref{fig:profexmpls}. Profiles of the Lyman lines of a given type (rows of
the figure) were not taken from the same position nor at the same time. On the
other hand, profiles of \ion{Mg}{ii} for each of the profile types are from the
same place and from the same time.

Comparing the all\discretionary{-}{-}{-}peak and above\discretionary{-}{-}{-}error maps 
shown in the left and right panels of Figs.~\ref{Fig:res_proftypemaps} and 
\ref{Fig:res_proftypemaps1}, one can conclude that peaks detected in profiles of the 
Ly$\beta$ and higher analyzed Lyman lines (maps for Ly$\gamma$ and Ly$\delta$ not shown, 
but are similar to those of Ly$\beta$) and in both \ion{Mg}{ii} k and h line profiles,
are often too small to be taken into account without any doubts caused by the noise. It 
means that they are not above error bars. This is also manifested in examples
of profiles in Fig.~\ref{fig:profexmpls}. Only for Ly$\alpha$, there are
a lot of
2\discretionary{-}{-}{-}peak profiles with both of them above error bars and
deep reversal in between them or profiles with more peaks but with two of them 
above error bars. This can be distinguished also on profile examples in the
left column of Fig.~\ref{fig:profexmpls}. These findings unambiguously lead to
more sophisticated classification, based on different combinations of counts of
peaks above error bars and peaks lost in noise.
For all six lines, the number of profiles with more than two peaks above error bars
in the analyzed area is much lower than the occurrence of profiles with one or two peaks
above error bars as it is prominent from the above\discretionary{-}{-}{-}error maps
shown in the right columns of Figs.~\ref{Fig:res_proftypemaps} and
\ref{Fig:res_proftypemaps1}. Moreover, the depth of central reversal and peak asymmetry
are defined only for 2\discretionary{-}{-}{-}peak profiles. Therefore, profiles with
more than two peaks
above error bars are not classified in any separate groups, instead they are all included
in the group containing peculiar profiles. Then, this naturally leads to classification
of profiles into the following six groups:
\begin{itemize}[labelwidth=5em,leftmargin=4.5em]
 \item[\hspace{5ex}1p] one peak
 \item[1p$_{\rm a}$\,\&\,mp$_{\rm b}$]  one peak above error bars (hereafter referred also
                                        to as peak above noise) and one or more additional
                                        peaks lost in noise (hereafter referred also to as
                                        subordinate peaks)
 \item[2p$_{\rm ab}$] two peaks, each of them above noise or subordinate
 \item[2p$_{\rm a}$]  two peaks, both above noise
 \item[2p$_{\rm a}$\,\&\,mp$_{\rm b}$] two peaks above noise plus one or more subordinate ones 
 \item[xp] peculiar profiles (with strange shapes without any clearly distinguishable peaks) 
           plus profiles not belonging to any of the five previous groups
\end{itemize}
In these groups, profiles are distinguished by the
self\discretionary{-}{-}{-}explaining designation encoding the number of
peaks (1, 2, or m meaning one or more) and their detectability – above error bars
indicated by the subscript $_{\mathrm{a}}$ or lost in noise (or below noise)
indicated by the subscript $_{\mathrm{b}}$. For instance, the group 2p$_{\rm ab}$
involves two\discretionary{-}{-}{-}peak profiles, each of the peaks can be
above error bars or lost in noise.
It should be emphasized that also here the xp group contains, in
addition to profiles with no distinguishable peaks, also profiles with more than
two peaks above error bars.
It was not necessary to develop any special method
for this classification, the profiles were classified according to already
detected peaks taking into account also their credibility.
The groups as they are defined, are not exclusive but their content can overlap.
This means that some profiles can belong to two groups at the same time as can be
seen in the diagrams
in Fig.~\ref{Fig:res_profgroupsplots}, where two different symbols representing different
groups can be located at the same position. We note that the profiles at the position
$[0,\,0]$ in the diagrams, are the peculiar profiles with strange shapes without any
clearly distinguishable peaks. One can notice
that there are remarkably larger portions
of such profiles in the \ion{Mg}{ii} lines. Relative counts in percents of profiles
classified in each of the six groups are also indicated in this figure. As was already
said, definition of the groups is not exclusive, therefore sum of counts of profiles 
in all groups as shown in each diagram in the figure can exceed $100$\,\%. Despite of
this, such definitions of groups are more suitable for later comparison with synthetic 
profiles produced by the non\discretionary{-}{-}{-}LTE modeling than classification
of profiles simply according to number of peaks marked regardless noise. 

In the case of Ly$\alpha$, there is $30$\,\% of 2\discretionary{-}{-}{-}peak 
profiles with both peaks above noise -- the group 2p$_{\rm a}$. If also profiles,  
with multiple peaks and two of them are above noise, are added -- the group 
2p$_{\rm a}$\,\&\,mp$_{\rm b}$, portion of profiles assumed as the 
2\discretionary{-}{-}{-}peak ones increases by $22$\,\% (up to more than $50$\,\%). 
For Ly$\beta$ and higher Lyman lines and the \ion{Mg}{ii} lines,  
there is much lower number of profiles in the group 2p$_{\rm a}$ ($12$\,\% for Ly$\beta$, 
and under $5$\,\% for Ly$\delta$ and the \ion{Mg}{ii} lines (even under $1$\,\% for 
\ion{Mg}{ii} h). Either contributions from the group 2p$_{\rm a}$\,\&\,mp$_{\rm b}$ is rather 
small for the Ly$\beta$\,--\,Ly$\delta$ lines. It is even under $1$\,\% for the \ion{Mg}{ii} 
lines, thus portion of profiles in the groups 2p$_{\rm a}$ plus 
2p$_{\rm a}$\,\&\,mp$_{\rm b}$ does not exceed $4$\,\%. Fortunately, there is much larger 
number of profiles in the IRIS spectral observations within the analyzed area than in the
SUMER spectral data ($11$ times more than in the SUMER spectral data in Ly$\alpha$ and $21$ 
times more than in the SUMER spectral data in Ly$\beta$\,--\,Ly$\delta$), thus, number 
of the \ion{Mg}{ii} k and h profiles belonging to groups 2p$_{\rm a}$ plus 
2p$_{\rm a}$\,\&\,mp$_{\rm b}$ is comparable to those of the Lyman lines. 
For the Ly$\alpha$\,--\,Ly$\gamma$ and \ion{Mg}{ii} k lines, portions of 
profiles with two peaks detected regardless noise (the group 2p$_{\rm ab}$) is between 
$20$ and $40$\,\%.  Only for the Ly$\delta$ and \ion{Mg}{ii} h lines, it is less 
($10$\,--\,$12$\,\%).  
%
%
\subsection{Statistics of the profile characteristics}
\label{Sec:results_profchars}
Histograms of integrated intensity ($E$) for profiles of all six spectral lines 
from analyzed area of the prominence are shown in Fig.~\ref{Fig:a1E} in the 
Appendix A. The histograms for three cases are plotted -- all profiles, profiles 
from the groups 1p plus 2p$_{\rm ab}$, and profiles from the groups 1p$_{\rm a}$\,\&\,mp$_{\rm b}$ 
plus 2p$_{\rm a}$\,\&\,mp$_{\rm b}$. The histograms for the 
Ly$\alpha$ have three maxima while shapes of the histograms for higher Lyman 
lines are simply single\discretionary{-}{-}{-}peaked. Histograms of $E$ for the 
\ion{Mg}{ii} k and h lines have similar shape with single peak as those for the 
Lyman lines and almost no differences occur between histograms for the three cases. 
Properties of the $E$ histograms for all six spectral lines for the three cases are 
listed in Table~\ref{tab:appb:EhistoProperties} in the Appendix B.  

In Fig.~\ref{Fig:a2C2P} in the Appendix A, histograms of the depth of the central 
reversal ($r_{\mathrm{CP}}$) for all six lines are plotted. The histograms for each 
line are plotted for three cases: profiles from the group 2p$_{\rm ab}$, group 
2p$_{\rm a}$\,\&\,mp$_{\rm b}$, and the combination of these two groups. Numbers of profiles 
considered as 2\discretionary{-}{-}{-}peak ones in each of the three cases, included in the 
histograms, are listed in Table~\ref{tab:appb:RCPhistoProperties} 
in the Appendix B as {$N$(2\discretionary{-}{-}{-}peak)}; numbers of profiles considered 
as one\discretionary{-}{-}{-}peaked --  not included in the histograms -- are denoted 
as {$N$(1\discretionary{-}{-}{-}peak)} in the table. For Ly$\alpha$, the
histogram for the first case contains the lowest count of profiles
({$N$(2\discretionary{-}{-}{-}peak)}),
while for other five lines, the lowest numbers of profiles is contained by the histogram 
for the second case. The histograms for the \ion{Mg}{ii} h and k lines for the second case 
contain even much lower counts of profiles (by one or even two orders of magnitude) than 
histograms for other two cases. The histograms for the all six lines and all three cases
have single peak. For Ly$\alpha$, the histograms for all three cases are almost the same.
On the other hand, the histograms for the higher Lyman lines for the second case differ
much from the histograms of other two cases. The histograms for Ly$\alpha$ and Ly$\beta$
are more\discretionary{-}{-}{-}or\discretionary{-}{-}{-}less symmetrical with clearly
distinguished maxima unlike those for Ly$\gamma$ and Ly$\delta$. Maxima of histograms
for the all four Lyman lines are at $r_{\mathrm{CP}}\sim0.25$\,--\,$0.50$.
The \ion{Mg}{ii} histograms for the second case have maxima at $\sim0.75$ while the
histograms for the first and third cases are strongly asymmetrical with maxima shifted
toward values of the $r_{\mathrm{CP}}$ ratio above $0.9$. Properties of the histograms
of $r_{\mathrm{CP}}$ for both Lyman and \ion{Mg}{ii} k and h lines together with number
of profiles assumed as 1\discretionary{-}{-}{-}peak and 2\discretionary{-}{-}{-}peaks
in each of the three cases, are listed in Table~\ref{tab:appb:RCPhistoProperties}
in the Appendix B.

The histograms of peak asymmetry ($r_{\mathrm{PA}}$) for all six spectral lines are
presented in the Appendix A in Fig.~\ref{Fig:a3PAs}. The
peak\discretionary{-}{-}{-}asymmetry histograms are also constructed for the same three
cases as it was for the $r_{\mathrm{CP}}$ histograms. The peak intensity ratios
(hereafter called as peak asymmetry) for each line and case are divided into two
histograms -- one for profiles with higher red peak than blue peak and another for
opposite asymmetry. Thus, the $r_{\mathrm{PA}}$ values range from zero up to unity in
each of the histograms. Only asymmetrical profiles (different heights of the peaks)
were taken into the histograms. Relative counts shown on ordinates were calculated
according to numbers of asymmetrical profiles (those with red peak dominating plus
those with blue peak dominating) which are indicated in
Table~\ref{tab:appb:RPAhistoProperties} as $N$\hbox{(2-peak~asm)}. Symmetric profiles
(with both peaks of equal height within a tolerance determined by size of one histogram
bin) are not included into histograms and their count is shown in the table as
$N$\hbox{(2-peak~sm)}. Profiles which peak asymmetry exceeds the tolerance are
assumed as asymmetrical. Number of 1\discretionary{-}{-}{-}peak
profiles, neither included into histograms, are denoted as $N$\hbox{(1-peak)}.
The histograms constructed for Ly$\alpha$ have typical shape with raise toward the
unity although this raise is not strictly monotonous and their maxima need not be at
their right edges. The histograms for the first and third cases are almost the same,
the histogram for the second case is little bit different. For higher Lyman lines,
histograms are gradually loosing this typical shape and more fluctuations also
occur.
Differences of the histograms for the second case from those for the first and third
cases, are also gradually getting larger. The histograms for the \ion{Mg}{ii} k and h
lines for all three cases resemble each other very much. They all are raising nicely
monotonously toward higher values of the $r_{\mathrm{PA}}$. Just the histograms for
the second case have some fluctuations, while the histograms for the other two cases
are nicely smooth. Thus, they reach their maxima at their right edges (toward the
unity), just the \ion{Mg}{ii} k\discretionary{-}{-}{-}line $r_{\mathrm{B-R}}$ histograms
for the first and third cases (group 2p$_{\mathrm{ab}}$\,, groups
2p$_{\mathrm{ab}}$ and/or 2p$_{\mathrm{a}}$\,\&\,mp$_{\mathrm{b}}$) have maxima shifted
left from right edges. For the k\discretionary{-}{-}{-}line $r_{\mathrm{B-R}}$
and h\discretionary{-}{-}{-}line $r_{\mathrm{R-B}}$ histograms, gradient of the raise
is not varying much through the whole histogram range. In difference from this, other 
\ion{Mg}{ii} histograms raise less steeply first and then, when approaching the unity,
the gradient suddenly increases.
%
%
\section{Discussion} 
\label{Sec:discussion}
\subsection{Profile types}
\label{sec:disc_proftype}
The first characteristics of observed spectral profiles we employ in this statistical analysis,  
is the shape of the profiles based on the number of distinct peaks (see Sect.~\ref{Sec:profpeaks}). 
Although this may seem as a trivial undertaking, we showed in Sect.~\ref{Sec:results_proftypes} that 
the classification of profiles based on the number of peaks is strongly influenced by the level of 
noise. This is true for both the Lyman lines and the \ion{Mg}{ii} k and h lines, and has several 
interesting consequences for eventual comparison with synthetic spectra produced by prominence 
radiative\discretionary{-}{-}{-}transfer models. The same classification based on the number of 
peaks, but without taking into account the effects of noise, was used for the \ion{Mg}{ii} k and 
h lines also by \citet{2018A&A...618A..88J} and \citet{2021A&A...653A...5P}.  

When we categorized the observed profiles of Lyman lines without taking into account the 
influence of noise, we found a large number of profiles that exhibit more than two
distinct peaks. An example of such multi\discretionary{-}{-}{-}peaked profiles is shown
in Fig.~\ref{fig:profexmpls}. These multi\discretionary{-}{-}{-}peaked profiles can be
a sign of highly dynamic behavior of unresolved prominence fine structures. Indeed,
dynamic fine structures can be seen in the peripheral parts of the studied prominence.
However, for the current statistical analysis, we used only the compact, dense and less
dynamic central part of the prominence (see Fig.~\ref{Fig:four_imgs_coal}). Moreover,
shapes of the broad Lyman lines are not remarkably influenced by the LOS velocities not
exceeding $10$\,--\,$20$\,km\,s$^{-1}$, typically present in quiescent prominences.
Another explanation for the multi\discretionary{-}{-}{-}peak Lyman profiles can be the
influence of noise. As we demonstrated in Fig.~\ref{Fig:res_proftypemaps}, that seems
to be the correct explanation. This is because when we take into account only those
peaks in the spectral profiles of the Lyman lines that extend above the noise level,
we obtain a majority of profiles classified as single or
double\discretionary{-}{-}{-}peaked. This result shows that the subordinate peaks can be
caused by noise while the dominant peaks are those that arise from the
non\discretionary{-}{-}{-}LTE radiative transfer effects. The same result also validates
the approach used in our previous statistical analysis of observed Lyman spectra of
prominences
\citep[][]{2010A&A...514A..43G,2015A&A...577A..92S}. In those works, we used 
double\discretionary{-}{-}{-}peaked profiles identified simply by their two dominant
peaks, without taking into account neither subordinate peaks nor the influence of the
noise.

Interestingly, while in the case of Lyman lines the introduction of the noise helped us
to identify most of the observed profiles as single or double-peaked, the situation for
the \ion{Mg}{ii} k and h lines is the opposite. The majority of the \ion{Mg}{ii} profiles
identified as double\discretionary{-}{-}{-}peaked
when we do not take the noise into account (see Fig.~\ref{Fig:res_proftypemaps1}) turned
out to have rather insignificant peaks that do not extend above the level of
noise -- examples of \ion{Mg}{ii} k and h profiles are shown in
Fig.~\ref{fig:profexmpls}. When taking the noise into account, these profiles are
classified as single-peaked or peculiar profiles due to their strong asymmetry and
generally flat tops. Again, broad and complex \ion{Mg}{ii} k and h profiles are a clear
sign of multiple dynamic fine structures present along a LOS
\citep[see][]{2020ApJ...888...42T,2022ApJ...934..133G}. However, in contrast to the
Lyman lines, the narrow \ion{Mg}{ii} k and h lines are highly sensitive to the LOS
velocities not exceeding $10$\,--\,$20$\,km\,s$^{-1}$. Therefore, the
hyper\discretionary{-}{-}{-}structuring of the \ion{Mg}{ii} k and h profiles analyzed
here is likely realistic and similar to that seen in the synthetic profiles of
\citet{2020ApJ...888...42T} or \citet{2022ApJ...934..133G}. Nevertheless,
in the observed spectra, the individual peaks cannot be conclusively discerned from
the random effects of noise.

Clearly, noise has a significant impact on the classification of spectral profiles using the number of 
peaks. This means that any method using such profile classification needs to pay a close attention to 
the fidelity of the peak detection. Due to the ambiguities in the detection of the peaks, it might  
also be worthwhile to use for comparison with synthetic spectra more robust profile characteristics 
that are less affected by noise. 
\subsection{Integrated intensities} 
\label{sec:disc_intint}
\begin{figure*}
\centering
\resizebox{0.95\hsize}{!}{
\includegraphics{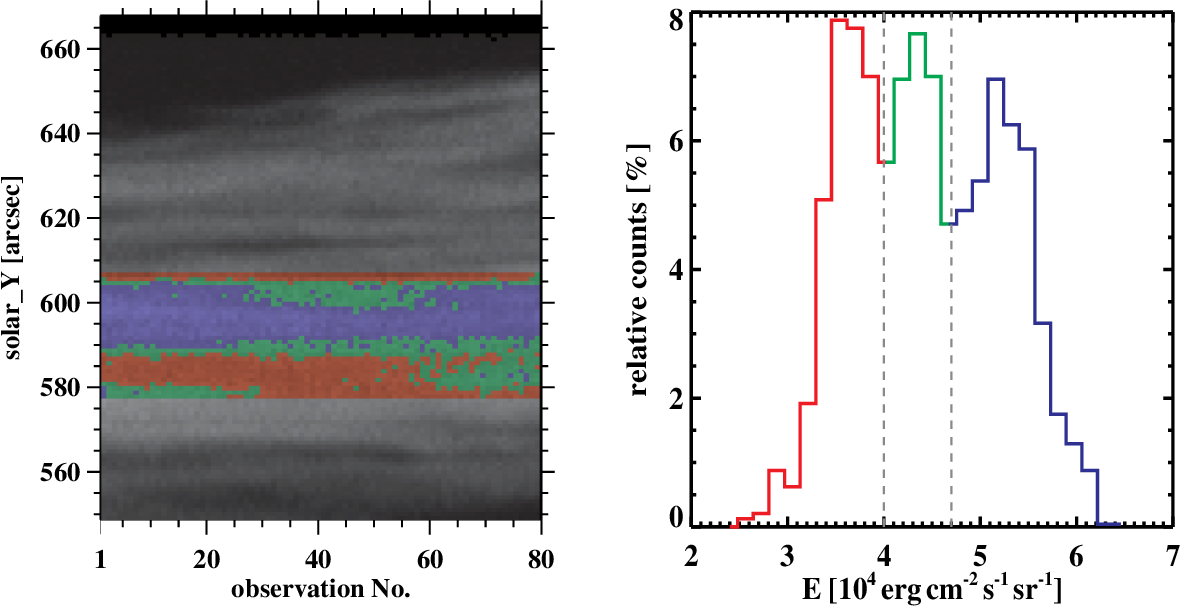}}
\caption{Map (left panel) showing the color mosaic of locations within the analyzed area 
with the Ly$\alpha$ integrated intensities belonging to the three peaks in the $E$ 
histogram (right panel). Integrated intensities outside of the analyzed area are 
shown in shades of gray. The histogram constructed for all Ly$\alpha$ profiles is 
used (shown also in the upper left panel of Fig.~\ref{Fig:a1E} in the Appendix A). Color 
coding in the mosaic corresponds to that used in the histogram: The red is for the left 
histogram peak, green for the middle, and blue for the right peak of the $E$ histogram.}
\label{Fig:trioehistly1}
\end{figure*} 
The second characteristics of the observed profiles that we study in the current work, is 
their integrated intensity $E$. To assess the influence of noise, we use here three sets of profiles. 
The first set contains all observed profiles, the second set contains 1\discretionary{-}{-}{-}peak 
and 2\discretionary{-}{-}{-}peak profiles identified without taking the noise into account (groups 
1p and 2p$_{\rm ab}$), and the last set contains only those 1\discretionary{-}{-}{-}peak and 
2\discretionary{-}{-}{-}peak profiles that have peaks extending above the level of noise 
(groups 1p$_{\rm a}$\,\&\,mp$_{\rm b}$ 
and 2p$_{\rm a}$\,\&\,mp$_{\rm b}$). As expected, noise has a negligible effect on the statistical properties of $E$ of all observed lines. This is demonstrated by the histograms shown in 
Fig.~\ref{Fig:a1E} in the Appendix A, which are nearly identical for all assumed sets of profiles. 

As we mentioned already, it is interesting that the histograms of $E$ of Ly$\alpha$ have three
distinct peaks while the histograms for higher Lyman lines and those of \ion{Mg}{ii} k and h 
lines are single\discretionary{-}{-}{-}peaked. In our previous statistical analyses of observed 
Lyman spectra, we obtained histograms of $E$ of all Lyman lines with a single peak 
(\citeauthor{2010A&A...514A..43G}\ \citeyear{2010A&A...514A..43G}) or with two peaks
(\citeauthor{2015A&A...577A..92S}\ \citeyear{2015A&A...577A..92S}). However, in the latter case, the 
histograms of all analyzed Lyman line profiles have two peaks and the second peak 
also appears in the histograms of $r_{\mathrm{CP}}$. In \citet{2015A&A...577A..92S} we thus could 
assume that the two\discretionary{-}{-}{-}peaked histograms indicate the presence of two independent 
prominence structures crossing the SUMER slit. 

In the case of the prominence studied here, only the $E$ histograms of Ly$\alpha$ have three
peaks. It is shown in Fig.~\ref{Fig:trioehistly1} that positions in the analyzed area of the
prominence with the Ly$\alpha$ $E$ va\-lu\-es belonging to each of the individual peaks are
gathered in several compact areas. For the figure, all Ly$\alpha$ profiles from the analyzed
area are taken and similar result were obtained when taking into account only profiles from
the groups $1\mathrm{p}$ and $2\mathrm{p}_{\mathrm{ab}}$, or
$1\mathrm{p}_{\mathrm{a}}\,\&\,\mathrm{mp}_{\mathrm{b}}$
and $2\mathrm{p}_{\mathrm{a}}\,\&\,\mathrm{mp}_{\mathrm{b}}$
(see Sect.~\ref{Sec:results_profchars}).
The $E$ intensities from the left histogram peak occur closer to edges of the analyzed area.
On the other hand, the highest $E$ values (the right histogram peak) occur only inside of
the analyzed area. Positions with $E$ values from the middle peak are located
mostly in few areas between those containing $E$ values from left and right peaks.
This indicates
that occurrence of three peaks in the $E$ histograms of Ly$\alpha$ is realistic, not caused by 
noise. Due to the fact that there are three peaks only in the $E$ histograms of Ly$\alpha$, 
it is clear that a simple explanation based on three distinct 
populations of prominence fine structures with different temperature and/or pressure 
(as it was in \citeauthor{2015A&A...577A..92S}\ \citeyear{2015A&A...577A..92S}) 
might not be sufficient. For such a case, the apparent trio of populations of Ly$\alpha$
profiles might help us to shed light on the composition of
the prominence\discretionary{-}{-}{-}to\discretionary{-}{-}{-}corona 
transition region (PCTR) surrounding the cool prominence material. This is because Ly$\alpha$ line 
is formed at higher temperatures closer to the surface of the observed prominence material than 
other Lyman lines \citep[][]{2005A&A...442..331H} or the \ion{Mg}{ii} k and h lines. Therefore, a 
difference in the plasma properties limited to the PCTR might lead to distinct populations only in 
the case of Ly$\alpha$. However, such an explanation will need to be supported by detailed 
radiative transfer modeling, which is planned for our next paper.
%
%
\subsection{Depths of the central reversal} 
\label{sec:disc_dor}
The third profile characteristics used in this work is the depth of the central reversal that describes 
the depth of the reversal in 2\discretionary{-}{-}{-}peak profiles. This ratio $r_{\mathrm{CP}}$ 
approaches to one when the profiles have shallow reversals and tends toward zero when the reversals are
extremely deep. 

From the description in Sect.~\ref{sec:disc_proftype}, it is clear that the identification 
of 2\discretionary{-}{-}{-}peak profiles is significantly influenced by noise. To assess this 
influence, we use two sets of profiles. The first set contains those profiles in which we have 
identified two peaks without taking the impact of noise into account (group 2p$_{\rm ab}$). 
The second set contains profiles in which two peaks extend above the level of the noise 
(group 2p$_{\rm a}$\,\&\,mp$_{\rm b}$). For calculations of the $r_{\mathrm{CP}}$ ratios
for profiles in this group, specific intensities at the two peaks above noise and reversal
in between are used. Thus, subordinate peaks are excluded from the calculations.

By taking into account the effect of noise, we selected 2\discretionary{-}{-}{-}peak profiles with 
sufficiently deep reversals, which results in histograms of $r_{\mathrm{CP}}$ shifted toward lower 
values (right upper panel and middle two panels of Fig.~\ref{Fig:a2C2P}). This is visible for higher 
Lyman lines (Ly$\beta$\,--\,Ly$\delta$). On the other hand, taking into account noise for Ly$\alpha$ 
has almost no effect on the histogram (upper left panel of Fig.~\ref{Fig:a2C2P}). All histograms 
of the depth of the central reversal for Lyman lines presented here are in strong contrast to the 
histograms for the observed data shown in \citet{2010A&A...514A..43G}. In fact, even without taking 
the noise into account (i.e., using the Lyman line profiles from group 2p$_{\rm ab}$), the histograms
for the prominence studied here show significantly deeper Lyman line profiles than those for the 
prominence studied in \citet{2010A&A...514A..43G}. This is interesting because the 2D 
multi\discretionary{-}{-}{-}thread non\discretionary{-}{-}{-}LTE radiative transfer 
models in the \citet{2010A&A...514A..43G} produced similarly deep synthetic profiles as those analyzed 
in the present work, but which were in stark contrast with the shallow observed profiles used in 
\citet{2010A&A...514A..43G}. A similar comparison can be done with histograms of the depth of 
the central reversal from \citet{2015A&A...577A..92S}. However, in this case, the situation is more 
complex because of the two distinct maxima in the histograms for observed data shown in that work. 
These two maxima likely correspond to two distinct populations of Lyman line profiles, one with deep 
reversals similar to the case presented here and the other with shallow reversals similar to those
of \citet{2010A&A...514A..43G}.

The statistics of $r_{\mathrm{CP}}$ becomes even more interesting 
in the case of the \ion{Mg}{ii} k and h lines (panels in the bottom row of Fig.~\ref{Fig:a2C2P}). 
Here, in the case when we do not take the effect of noise into account, we obtained histograms 
that strongly tend toward shallow profiles. On the other hand, when we select only those 
\ion{Mg}{ii} k and h profiles that have peaks extending above the noise level (group 
2p$_{\rm a}$\,\&\,mp$_{\rm b}$), we obtained $r_{\mathrm{CP}}$ histograms with median values 
around $0.7$\/. This should not come as a surprise because we have basically selected only a 
small portion (around $6$\,\% for \ion{Mg}{ii} k) of the observed profiles that have deep
reversals. However, the strong difference between the resulting histograms shows the 
importance of the correct choice of statistical characteristics for any comparison with 
synthetic profiles. 

Histograms constructed for combination of the two groups -- 2p$_{\rm ab}$ and/or
2p$_{\rm a}$\,\&\,mp$_{\rm b}$ -- are similar to those constructed for not taking noise 
into account; for the \ion{Mg}{ii} lines these histograms are almost identical.  
This indicates larger number of the 2p$_{\rm ab}$ profiles than those from the 
2p$_{\rm a}$\,\&\,mp$_{\rm b}$ group (see Table~\ref{tab:appb:RCPhistoProperties} in the 
Appendix B). Thus, the 2p$_{\rm ab}$ profiles dominate in determining shapes of the histograms. 

The main role in the distinction between a large number of shallow profiles 
and a small number of profiles with deep reversals, which manifests mainly for the \ion{Mg}{ii} 
k and h lines, is played by the dynamics of multiple fine structures 
along a LOS. The shallow profiles are observed in places where multiple fine structures with different 
LOS velocities are crossing the LOS while the deep profiles likely correspond to instances when
we observe single fine structures or more fine structures but with similar LOS velocities
\citep[see][for more details]{2020ApJ...888...42T,2022ApJ...934..133G}. This means that when we are 
interested in the diagnostics of the dynamics of prominence fine structures, we need to use 
the $r_{\mathrm{CP}}$ histograms containing all 2\discretionary{-}{-}{-}peak \ion{Mg}{ii} k and h 
profiles regardless of noise. On the other hand, if we want to diagnose the temperature and 
pressure properties of the prominence fine structures, we need to focus on the deep profiles that 
are minimally affected by the dynamics. 
\subsection{Peak asymmetry}
\label{sec:disc_pas}
The last profile characteristics considered in this work is the asymmetry of the peaks 
$r_{\mathrm{PA}}$ of 2\discretionary{-}{-}{-}peak profiles. This proved to be helpful
for the diagnostics of the LOS dynamics of prominence fine structures
in \citet{2010A&A...514A..43G} and was considered also by \citet{2018A&A...618A..88J}
and \citet{2021A&A...653A...5P} for the
\ion{Mg}{ii} k and h lines. Since the identification of the peaks is affected by noise, we again 
use two sets of profiles to assess its influence. The first set contains those profiles in which 
we have identified two peaks without taking the impact of noise into account (group 2p$_{\rm ab}$) 
while the second set contains profiles in which two peaks extend above the level of the noise 
(group 2p$_{\rm a}$\,\&\,mp$_{\rm b}$). 
Also here, similarly as for the depth of the central reversal, the asymmetry of the peaks
is calculated using specific intensities only at the two peaks above noise (subordinate profiles 
are excluded from the calculations). In the case of the Lyman lines, the histograms of peak
asymmetries in both cases -- without or with taking noise into account -- seem similar (the
first part of Fig.~\ref{Fig:a3PAs} in the Appendix A). Just for more symmetrical profiles
(values of $r_{\mathrm{PA}}$ above $0.8$), there are larger relative counts in the histograms 
for the latter case. It can be explained by the fact that the profiles with deeper reversals 
occurring mostly in the 2p$_{\rm a}$\,\&\,mp$_{\rm b}$ group (as already stated in 
Sect.~\ref{sec:disc_dor}), are less sensitive to dynamics of the prominence plasma due to larger 
optical thickness in cores of these profiles. All histograms presented here are rather different 
from those obtained in \citet{2010A&A...514A..43G}. In that work, all histograms were monotonously 
raising toward unity. But some of the histogram presented here have their global maxima at lower 
values of $r_{\mathrm{PA}}$. This is most probably caused by rather large dynamics 
of the prominence plasma. 

Although histograms of the \ion{Mg}{ii} k and h lines (second part of Fig.~\ref{Fig:a3PAs} in the 
Appendix A) for the two sets of profiles are very similar, the application of noise leads to 
significantly smaller sample of profiles used for the statistics (see Table 
\ref{tab:appb:RPAhistoProperties} in the Appendix B). This causes that the histograms for 
profiles from the 2p$_{\rm a}$\,\&\,mp$_{\rm b}$ group are less smooth than those for the 
group 2p$_{\rm ab}$. These selected profiles (those with peaks extending above the level of noise) 
also tend to be more symmetric, having the ratio of one peak to another nearing the unity. This 
is not surprising. As we discuss in Sect.~\ref{sec:disc_dor}, the \ion{Mg}{ii} k and h profiles 
with extended peaks likely correspond to arrangements of fine structures with limited LOS dynamics. 
Such configurations then result in small peak asymmetries. Therefore, when we want to diagnose the 
dynamics of prominence fine structures (for which $r_{\mathrm{PA}}$ is a suitable parameter), 
we need to use all 2\discretionary{-}{-}{-}peak profiles for the statistical comparison with the 
synthetic spectra. 
%
%
\subsection{Possible biases}
\label{sec:possbiass} 
Conclusions drawn in this study may be biased by potentially short exposure time of the IRIS 
\ion{Mg}{ii} spectra, large imbalance of statistical samples of Lyman and \ion{Mg}{ii} spectra, 
and by an ad hoc choice of bin size
in the histograms. The first bias
manifests itself by the noise resulting in an ambiguity in peak detection even in the brightest 
region of prominence. The issue can be tackled in the future by an
trial\discretionary{-}{-}{-}and\discretionary{-}{-}{-}error finding of
an optimum exposure time for prominence \ion{Mg}{ii} spectra. The second bias is apparent if we 
realize an order of difference between the number of profiles (samples) employed in plotting 
the diagrams for the Lyman and \ion{Mg}{ii} spectra (Fig.~\ref{Fig:res_profgroupsplots}). 
Obviously, this is given by different spatial sampling and duration of SUMER and 
IRIS spectral data. The pixel sizes along the SUMER and IRIS slits are 
$1$\,arcsec (Sect.~\ref{Sec:sumer}) and $0.166$\,arcsec (Sect.~\ref{Sec:iris}), respectively.
While the SUMER observations in one spectral lines
lasted less than $20$\,min (Table~\ref{tab:SumObs}), the IRIS observations
lasted $39$\,min (Sect.~\ref{Sec:iris}). The third bias may influence the appearance
of histograms in the Appendix~A but should not change much
the numerical values in the Appendix~B. The issue may be mitigated
by employing large statistical samples for which histograms are likely to be less
sensitive to bin size.
%
%
\section{Conclusions}
\label{Sec:conclusions}
In this work, we present a statistical spectroscopic analysis of a unique dataset of coordinated prominence observations in the Lyman lines
(Ly$\alpha$ to Ly$\delta$) and the \ion{Mg}{ii} k and h lines. The observed data were obtained quasi co\discretionary{-}{-}{-}spatially and almost
simultaneously by two space spectrographs -- SoHO/SUMER and IRIS. Only a few similar coordinated datasets of Lyman and \ion{Mg}{ii} k and h
observations have ever been obtained in prominences and we present here the first analysis using these two sets of spectral lines. Such broad sets of
spectroscopic observations in multiple lines represent an excellent opportunity for diagnostics of the properties of the prominence plasma and the
dynamics of their fine structures. In the current work, we studied in detail the following profile characteristics: the shape of the observed line
profiles based on the number of distinct peaks, the integrated line intensity, the depth of the central reversal of 
two\discretionary{-}{-}{-}peaked profiles, and the asymmetry of these peaks. Moreover, for the first time, 
we assessed the influence of noise on the statistical properties of these profile characteristics. 

As expected, noise has a negligible effect on $E$ of all observed lines. However, the presence of noise significantly
affects the classification of spectral profiles using the number of peaks, $r_{\mathrm{CP}}$, and also 
$r_{\mathrm{PA}}$. By taking the influence of noise into account, we could demonstrate which profile 
characteristics in which spectral lines are suitable for diagnostics of different properties of the 
observed prominence. 

In the case of the Lyman lines, we showed that subordinate peaks are mostly caused by noise 
while the dominant peaks are those that arise from the non\discretionary{-}{-}{-}LTE radiative 
transfer effects. This is consistent with the fact that Lyman lines are not strongly sensitive 
to the typical LOS velocities
($10$\,--\,$20$ \,km\,s$^{-1}$) exhibited by the fine structures of quiescent prominences.
In Lyman lines, such velocities are manifested by the asymmetry of the dominant peaks
\citep[see, e.g.,][]{2008A&A...490..307G,2010A&A...514A..43G},
but they do not cause multiple peaks 
\citep[unlike in the case of the \ion{Mg}{ii} k and h lines, see, e.g.,][]{2015ApJ...800L..13H}.
Moreover, the LOS velocities on the order of $10$\,km\,s$^{-1}$ were found to be present in the
analyzed prominence by \citet{2018A&A...618A..88J}. Therefore, noise is the most likely 
cause of the subordinate peaks in the analyzed Lyman line profiles. This, however, means that 
only the dominant peaks of the reversed profiles of Lyman lines should be used for statistical 
analyses or comparisons with synthetic spectra 
\citep[as in, e.g.,][]{2010A&A...514A..43G,2015A&A...577A..92S}. The classification of Lyman line
profiles based on an overly sensitive identification of peaks can lead to many unnecessarily 
discarded observed profiles which may significantly degrade the statistical significance of 
observed datasets. The misclassification of profiles as multi\discretionary{-}{-}{-}peaked
can also make the fitting
of observed and synthetic profiles (which do not suffer from noise) unreliable. These findings mean 
that any method for the identification of peaks in the observed Lyman line profiles needs to pay 
close attention to the impact of noise. 

The situation is different in the case of the \ion{Mg}{ii} k and h lines. As we showed, 
the majority of the \ion{Mg}{ii} k and h profiles identified as double-peaked when we do not take 
the noise into account have in fact rather insignificant peaks that do not extend above the level 
of noise. Therefore, due to the presence of noise, we are unable to unequivocally identify peaks in 
many \ion{Mg}{ii} k and h profiles. Often there are simply too many low peaks. However, based on the 
fact that the \ion{Mg}{ii} k and h lines are highly sensitive to the LOS velocities on the order of 
$10$\,-\,$20$\,km\,s$^{-1}$, this hyper\discretionary{-}{-}{-}structuring
present in the \ion{Mg}{ii} k and h profiles is likely
realistic and similar to that seen in the synthetic profiles of \citet{2020ApJ...888...42T}
or \citet{2022ApJ...934..133G}. This means that by taking the impact of noise into account,
we effectively selected only those \ion{Mg}{ii} k and h profiles that have sufficiently
high peaks and thus a deep reversal. Such deeply reversed profiles may represent only a
small fraction of the observed profiles,
but they nevertheless provide a useful diagnostics tool. The key role in the distinction between 
the multi\discretionary{-}{-}{-}peaked \ion{Mg}{ii} k and h profiles with low peaks and the profiles 
with deep reversals is played by the dynamics of multiple fine structures located along a LOS. 
The complex, multi\discretionary{-}{-}{-}peaked profiles are observed
in places where multiple fine structures with different
LOS velocities are crossing the LOS, while the profiles with deep reversals likely correspond to 
instances when we observe single fine structures or more fine structures but with similar LOS 
velocities \citep[see][for more details]{2020ApJ...888...42T,2022ApJ...934..133G}. This means that 
when we are interested in the diagnostics of the dynamics of prominence fine structures, we need to 
use all \ion{Mg}{ii} k and h profiles. On the other hand, if we want to diagnose the temperature 
and pressure properties of individual prominence fine structures, we need to focus on the deep 
profiles that are minimally affected by the fine-structure dynamics. 

Another profile characteristic useful for the diagnostics of dynamics is the profile asymmetry. 
However, in this case, the \ion{Mg}{ii} k and h profiles with deep reversals are not very useful 
because they tend to be more symmetric. This is not surprising. As we discussed above, such
\ion{Mg}{ii} k and h profiles likely correspond to arrangements of fine structures with
limited LOS dynamics, resulting in small $r_{\mathrm{PA}}$ values. The asymmetry of the peaks
in the Lyman lines (when the
influence of noise is taken into account) is more suitable for the analysis of the prominence 
fine\discretionary{-}{-}{-}structure dynamics. We want to note here that although the era of 
SoHO/SUMER observations is over,
the Solar\discretionary{-}{-}{-}C mission\footnote{\url{https://solar-c.nao.ac.jp/en/}}
(\citeauthor{10.1117/12.2560887}\ \citeyear{10.1117/12.2560887}, and especially
\citeauthor{10.1117/12.2599610}\ \citeyear{10.1117/12.2599610}, presenting a proposal
of instrumental design of the EUV High\discretionary{-}{-}{-}Throughput Spectroscopic
Telescope  planed for  Solar\discretionary{-}{-}{-}C) will provide new Lyman line
observations in the future.

To conclude, the best approach to the analysis of the prominence fine structure dynamics is 
to use a combination of profile asymmetry in Lyman lines together with complex profiles of \ion{Mg}{ii} 
k and h lines. On the other hand, for diagnostics of the prominence plasma parameters, such as 
temperature or pressure, the best thing is to concentrate on the deeply reversed \ion{Mg}{ii} k and h lines 
in combination with Lyman lines and analyze the depth of the central reversal and the integrated 
intensities. 

This study highlights the complexities we encounter when we want to compare the observed 
spectra with the synthetic spectra provided by models. Previously, three different approaches were 
used: i) direct fitting of the shape of the line profiles, ii) fitting of selected line parameters, 
and iii) comparison of statistical properties of selected profile characteristics. 

The first approach relies on detailed matching of the actual shapes of observed profiles. 
Examples of this approach are the ad\discretionary{-}{-}{-}hoc fitting of the Lyman line profiles
done by \citet{2007A&A...472..929G} or the automated \ion{Mg}{ii} k and h profile fitting
procedures of \citet{2021A&A...653A..94B} and \citet{2021A&A...653A...5P}. However, as we
demonstrate in this work, the
shape of the observed profiles is strongly influenced by the presence of noise. This means that 
special attention needs to be paid to the classification of the observed spectra. Moreover, as 
we have discussed here, the \ion{Mg}{ii} k and h lines are highly sensitive to the fine-structure 
dynamics, which results in complex, multi\discretionary{-}{-}{-}peaked profiles. Therefore, the 
best set of \ion{Mg}{ii} k and h profiles suitable for the direct 
profile\discretionary{-}{-}{-}to\discretionary{-}{-}{-}profile comparison 
is the deeply reversed profiles with peaks unequivocally extending above the level of noise. 
As shown by \citet{2021A&A...653A..94B} and \citet{2021A&A...653A...5P}, such \ion{Mg}{ii} k 
and h profiles can be matched by synthetic profiles with a good confidence. However, these 
deeply reversed profiles are typically present only in the outskirts of prominences 
\citep[see][]{2021A&A...653A..94B,2021A&A...653A...5P},
where the fine\discretionary{-}{-}{-}structure dynamics plays
a marginal role. There, the direct profile\discretionary{-}{-}{-}to\discretionary{-}{-}{-}profile 
fitting provides good diagnostics of the prominence plasma parameters, such as temperature and 
pressure. 

The second approach is the fitting of profile parameters,
such as in the non\discretionary{-}{-}{-}LTE inversion
method of \citet{2022ApJ...932....3J}. This method allows the use of profile parameters that are 
not strongly affected by noise (such as integrated intensities); however, it may not lead to a
unique solution, especially in the case of complex observed profiles
\citep[see][for more details]{2022ApJ...932....3J}. 

The third approach does not focus on the individual line profiles but compares the 
statistical properties of selected profile characteristics of entire observed datasets.
Examples of this approach are
\citet{2010A&A...514A..43G} and \citet{2015A&A...577A..92S}. The statistical comparison is
well suited for analysis of the overall properties of observed prominences, such as the general 
level of the fine\discretionary{-}{-}{-}structure dynamics. However, as we have shown
in the present work, due to the impact of noise, close attention needs to be paid to the
selection of the profile characteristics used for the comparison.

As we can see, broad datasets comprising many different spectral lines represent a great 
diagnostic potential but also significant challenges for comparison with synthetic spectra.
These challenges are even greater in the case of spectral lines sensitive to the
fine\discretionary{-}{-}{-}structure dynamics, such as the \ion{Mg}{ii} k and h lines.
Therefore, for future comparison with the results of complex
multi\discretionary{-}{-}{}dimensional models with dynamics fine structures, we may
need to combine these three approaches. To be less sensitive to the impact of noise
in the line profiles without having to reduce it, it could be useful to introduce
other profile characteristics that are robust with respect to noise (such as the
integrated intensity -- see Sect.~\ref{Sec:discussion}, or the first and second
moments of profiles), but that are still capable of describing
multi\discretionary{-}{-}{-}peak profiles. However, solely using such
characteristics may lead to an unwitting introduction of biases. On the other
hand, spectroscopic analysis based on the categorization of profiles according to
the number of peaks or made using the characteristics for
multi\discretionary{-}{-}{-}peak profiles, which are defined for
two\discretionary{-}{-}{-}peak profiles (i.e., two peaks are selected only
and others are neglected), can be highly sensitive to the presence of noise.
However, selecting peaks above noise only and assuming others as noise might not be
correct because some of the subordinate peaks could be real peaks and their
occurrence can bear important information on the dynamics of the investigated object.
Thus, to distinguish between the real subordinate peak and
noise correctly, one has
to keep in mind the properties of the observed spectral lines and the dynamics of the observed
object (i.e., optical thickness of observed lines and LOS velocities). For the analysis of
the quiescent prominence observed in the Ly$\alpha$\,--\,Ly$\delta$
and \ion{Mg}{ii} k and h lines carried out here, as already stated, the LOS velocities
should not be large enough so as to cause any multi\discretionary{-}{-}{-}peak
Lyman\discretionary{-}{-}{-}line
profiles due to a rather large optical thickness of prominence plasma in the cores of these lines. Conversely, for the less optically thick \ion{Mg}{ii} lines,
multi\discretionary{-}{-}{-}peak profiles can also arise for such LOS velocities.
In the future, a mitigation of noise using deconvolution techniques removing
instrumental effects including noise before the analysis
(see, e.g., \citeauthor{2023A&A...679A.156P} \citeyear{2023A&A...679A.156P})
will be more effective than the subsequent distinguishing of real peaks from noise.
These numerical techniques will continue to improve
and conclusions of our present a\-na\-ly\-sis could be an impulse for it.
It has to be noted that setting an optimum exposure time
can also be important for lowering noise in observed spectroscopic data.
%
%
%
%
\begin{acknowledgements}
P.S. and J.K. acknowledge support from the project VEGA 2/0043/24 of the Science Agency.
S.G., P.H., J.K., and P.S. acknowledge support from the grant No.~22-34841S of the 
Czech Science Foundation (GA\v{C}R). S.G., P.S., P.H., and J.K. acknowledge support 
from the Joint Mobility Project SAV\discretionary{-}{-}{-}18\discretionary{-}{-}{-}03 
of the Academy of Sciences of the Czech Republic and Slovak Academy of Sciences. 
S.G. and P.H. thank for support from the project RVO:67985815 of the Astronomical 
Institute of the Czech Academy of Sciences. 
P.H. acknowledges support by the program ’Excellence Initiative - Research University’ 
for years 2020-2026 at University of Wroclaw, project No.~BPIDUB.4610.96.2021.KG. 
Authors are thankful to Dr. W. Curdt for his useful information and advices according 
to SUMER observations used here. IRIS is a NASA small explorer mission developed and 
operated by LMSAL with mission operations executed at NASA Ames Research Center and 
major contributions to downlink communications funded by ESA and the Norwegian Space 
Center. This research has made use of NASA’s Astrophysics Data System. We are thankful 
to the anonymous referee for his/her useful comments and suggestions which improved 
this publication much. 
\end{acknowledgements}


%
%
\begin{appendix}
\onecolumn
\addtolength{\topmargin}{0.3cm}
\addtolength{\topskip}{0.5cm}
\section{Histograms of the profile characteristics of hydrogen Lyman 
and \ion{Mg}{ii} h and k lines}
\begin{figure}[H]
\centering
\resizebox{\hsize}{!}{
\includegraphics{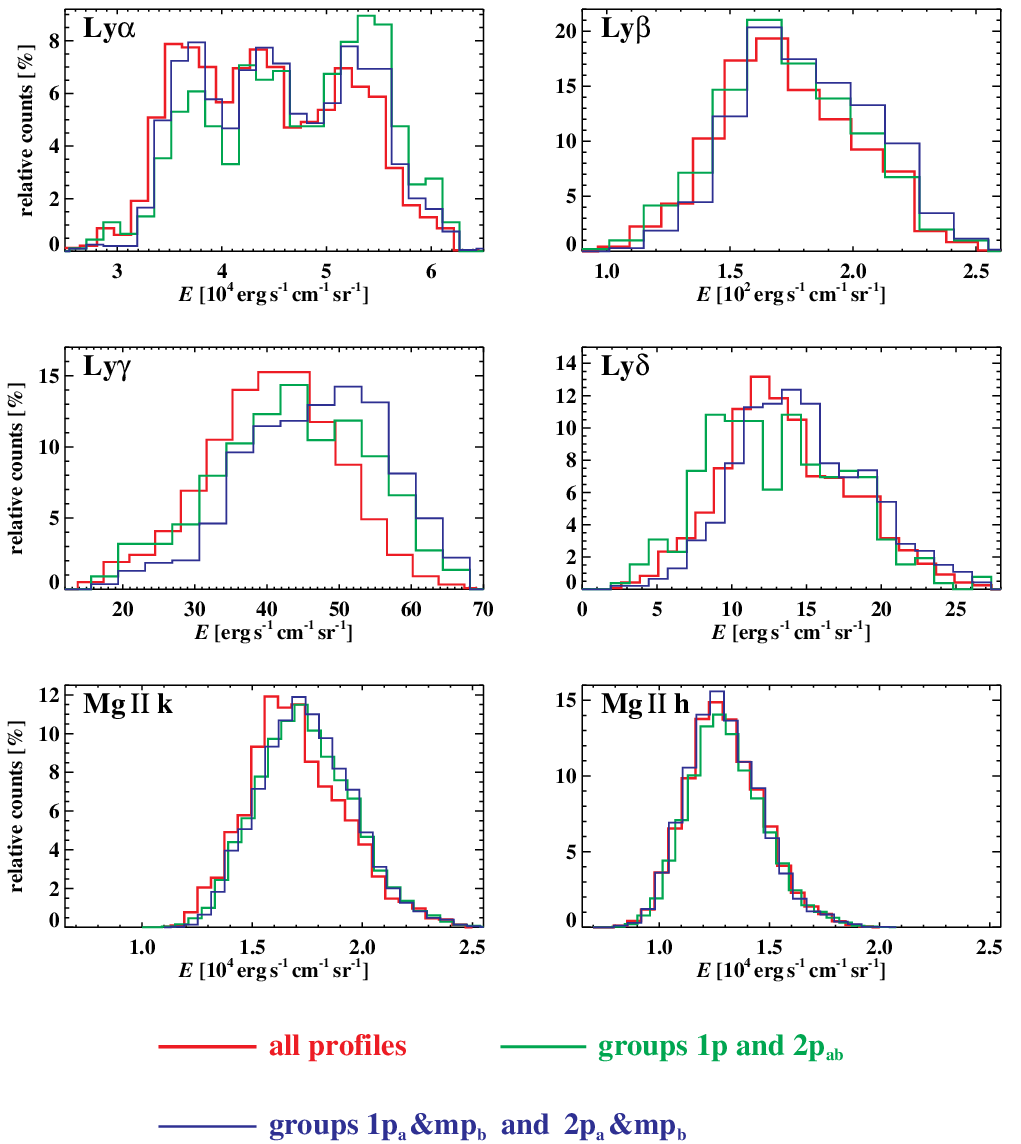}} 
\caption{Histograms of the integrated intensities of the Ly$\alpha$\,--\,Ly$\delta$ and 
\ion{Mg}{ii} k and h lines. Histograms containing all types of profiles are plotted 
with red, profiles from combination of the groups 1p and 2p$_{\rm ab}$ with green, 
and from combination of the groups 1p$_{\rm a}$\,\&\,mp$_{\rm b}$ and 
2p$_{\rm a}$\,\&\,mp$_{\rm b}$ with the blue color.} 
\label{Fig:a1E}
\end{figure}
\addtolength{\topmargin}{-0.3cm}
\addtolength{\topskip}{-0.5cm}
\begin{figure}[H]
\centering
\resizebox{\hsize}{!}{
\includegraphics{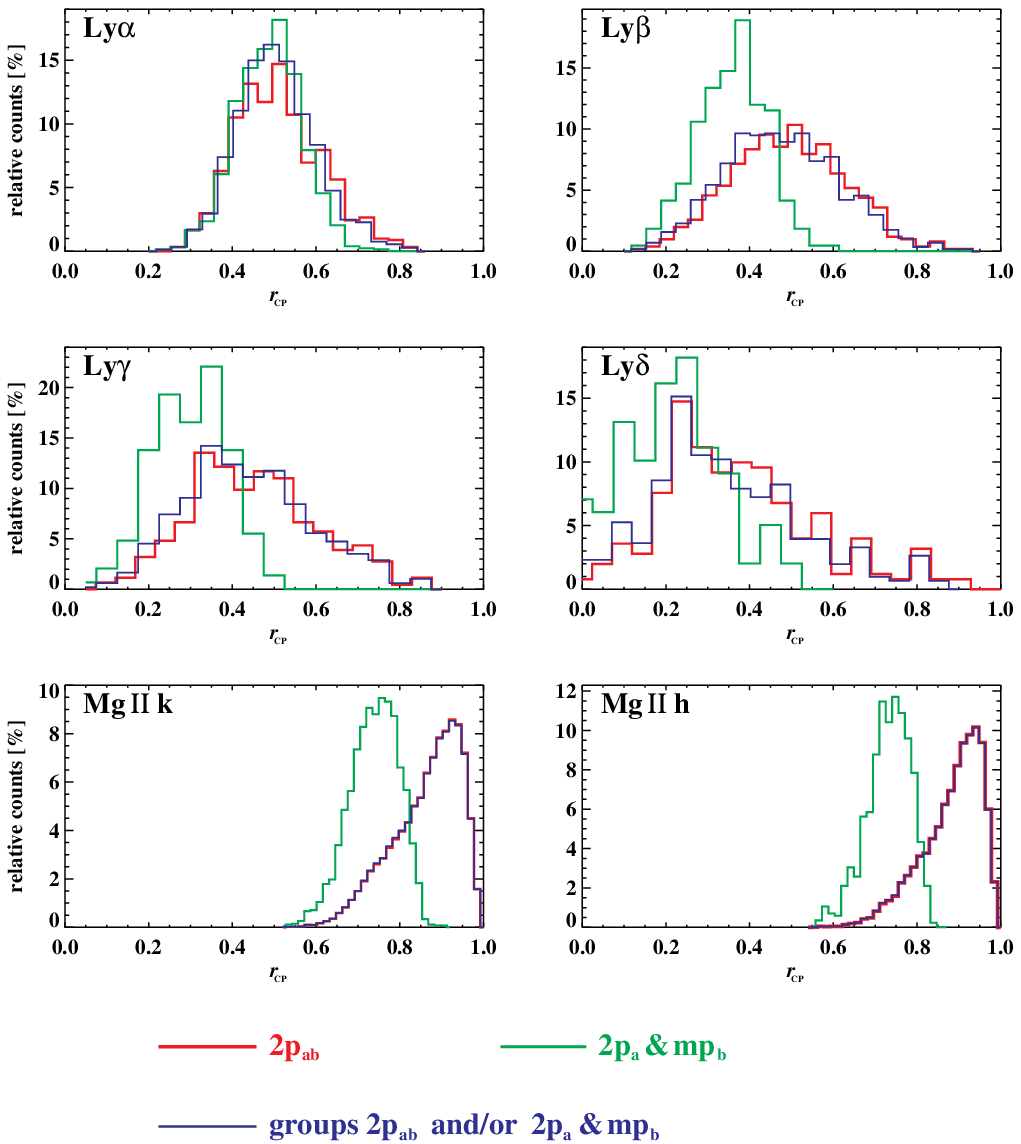}} 
\caption{Histograms of the depth of the central reversal of the Ly$\alpha$\,--\,Ly$\delta$ and 
\ion{Mg}{ii} k and h lines. Histogram for profiles from the group 2p$_{\rm ab}$ are plotted 
with red, for profiles from the group 2p$_{\rm a}$\,\&\,mp$_{\rm b}$ with green, and for profiles
occurring in combination of these groups (2p$_{\rm ab}$ and/or 2p$_{\rm a}$\,\&\,mp$_{\rm b}$) with 
the blue line.} 
\label{Fig:a2C2P}
\end{figure}
\begin{figure}[H]
\centering
\resizebox{\hsize}{!}{
\includegraphics{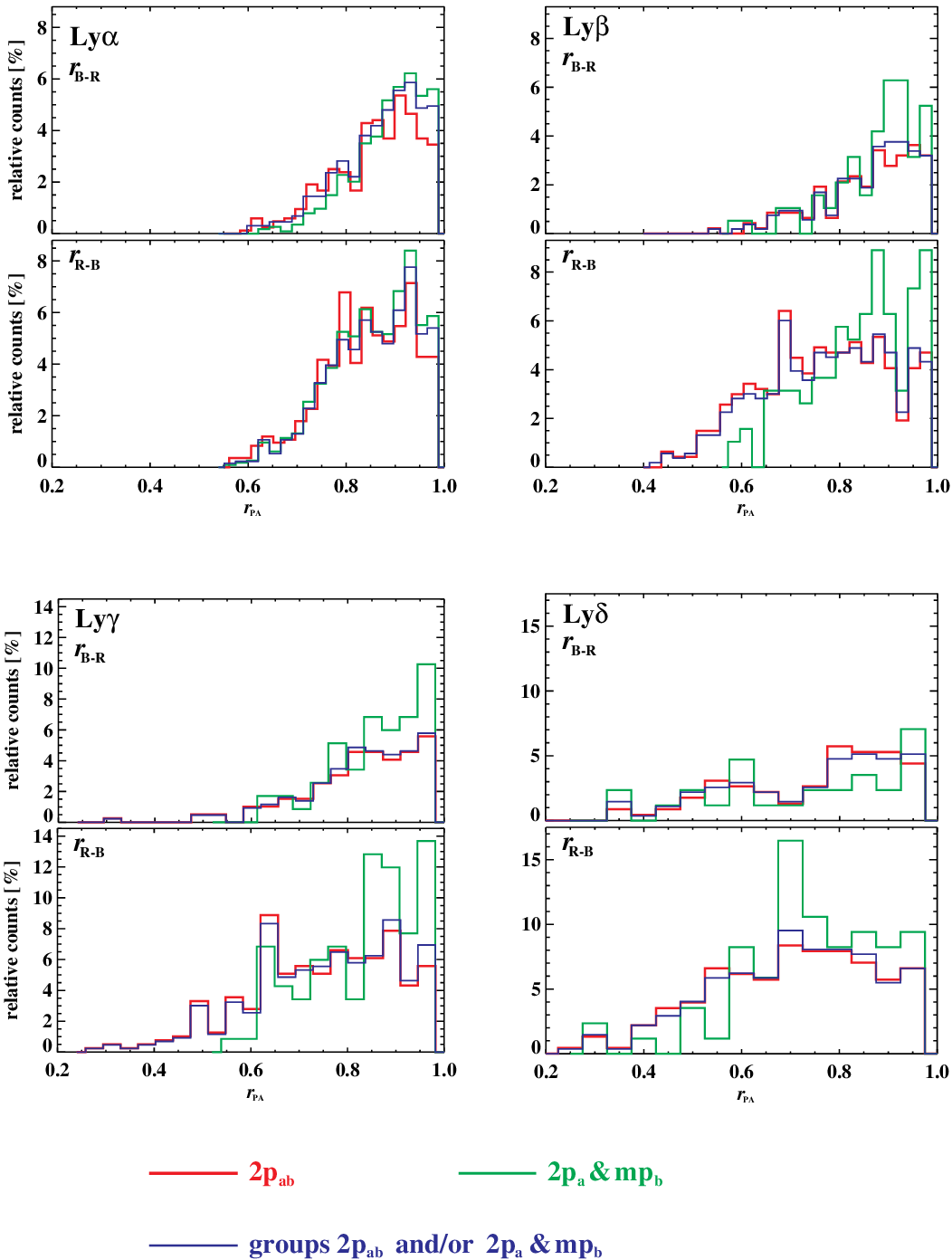}} 
\caption{To be continued on the next page}
\label{Fig:a3PAs}
\addtocounter{figure}{-1}
\end{figure}
\begin{figure}[H]
\centering
\resizebox{\hsize}{!}{
\includegraphics{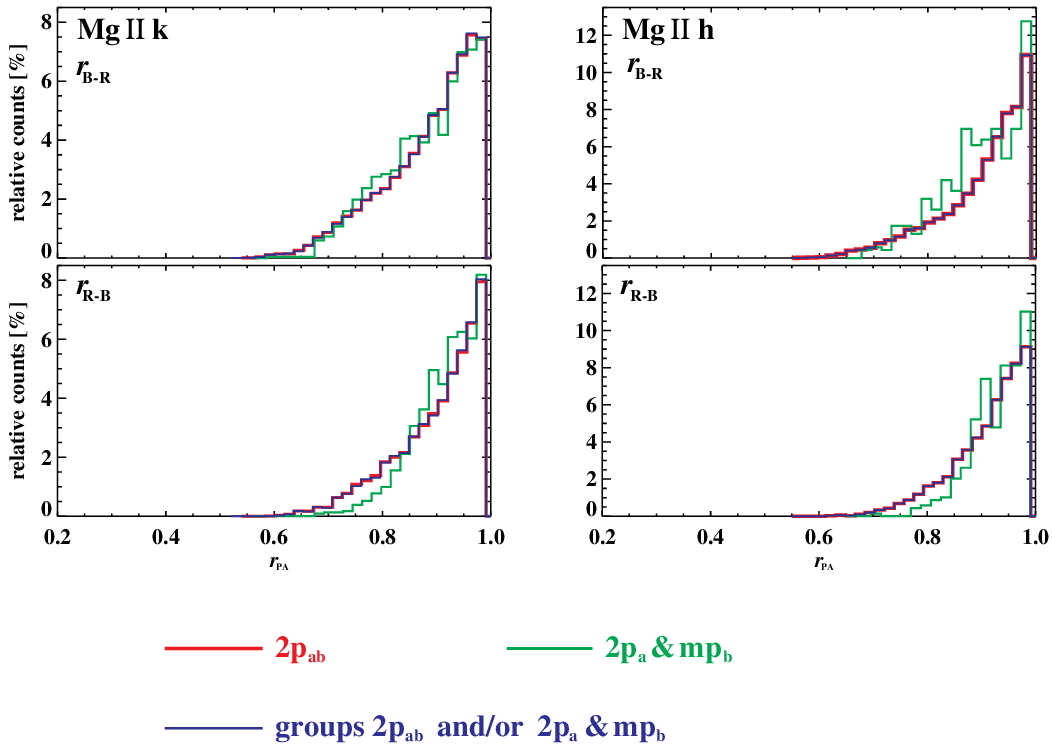}} 
\caption{Histograms of the peak asymmetry ($r_{\mathrm{PA}}$) of the 
Ly$\alpha$\,--\,Ly$\delta$ and \ion{Mg}{ii} k and h lines. Individual histograms are constructed 
for profiles with dominating red peak (denoted with label $r_{\mathrm{B-R}}$) and dominating 
blue peak ($r_{\mathrm{R-B}}$). For all six lines, histograms for three cases are plotted with 
lines of different colors: Those constructed for profiles from the group 2p$_{\rm ab}$ 
with red, for profiles from the group 2p$_{\rm a}$\,\&\,mp$_{\rm b}$ with green line. For a
combination of profiles from both these groups, the histograms are plotted with the blue line.}
\end{figure}
%
%
%
%
\newpage
\addtolength{\topmargin}{0.3cm}
\addtolength{\topskip}{0.5cm}
\section{Properties of histograms of the profile characteristics of hydrogen Lyman 
and \ion{Mg}{ii} h and k line profiles}
%
%
\begin{table}[H]
\centering
\caption{Properties of histograms of the integrated intensity $E$ shown in Fig.~\ref{Fig:a1E} 
in Appendix A for the six studied lines and three cases: 1\,--\,all profiles,
2\,--\,profiles from the groups 1p and 2p$_{\rm ab}$\hspace{.2ex}, and
3\,--\,profiles from the groups 1p$_{\rm a}$\,\&\,mp$_{\rm b}$ and
2p$_{\rm a}$\,\&\,mp$_{\rm b}$.}
\label{tab:appb:EhistoProperties}
\begin{tabular}{cccccr}
\hline
\hline
     &      &        &              &           &            \\[-2.2ex]
line & case\hspace{3ex} & median & med. of dev. & hist. max & $N$ profiles  \\
     &    &   $\left[\mathrm{erg}\,\mathrm{s}^{-1\ }\mathrm{cm}^{-2\ }\mathrm{sr}^{-1}\right]$  & 
     $\left[\mathrm{erg}\,\mathrm{s}^{-1\ }\mathrm{cm}^{-2\ }\mathrm{sr}^{-1}\right]$ 
    & $\left[\mathrm{erg}\,\mathrm{s}^{-1\ }\mathrm{cm}^{-2\ }\mathrm{sr}^{-1}\right]$   &  \\[0.8ex] \hline
     &      &              &              &             &         \\[-2.2ex]
     &  1   &  $4.5$E$+04$ &  $6.3$E$+03$ & $3.5$E$+04$ &  $2400$ \\[-0.7ex]
 Ly$\alpha$ &  2   & $4.8$E$+04$ &  $5.8$E$+03$ & 
          $5.4$E$+04$ &  $905$  \\[-0.7ex]
    &  3   & $4.6$E$+04$ &  $6.4$E$+03$ & 
          $3.8$E$+04$ &  $1990$ \\[0.8ex]
     &  1   & $1.8$E$+02$ &  $1.8$E$+01$ & 
          $1.7$E$+02$ & $1200$ \\[-0.7ex]
 Ly$\beta$ &  2   & $1.8$E$+02$ &  $2.0$E$+01$ & 
          $1.6$E$+02$ & $504$  \\[-0.7ex]
           &  3   & $1.9$E$+02$ &  $1.9$E$+01$ & 
          $1.6$E$+02$ & $693$  \\[0.8ex]
     &  1   & $4.3$E$+01$ &  $6.0$E$+00$ & 
        $4.1$E$+01$ & $1200$  \\[-0.7ex]
Ly$\gamma$ &  2   &  $4.6$E$+01$ &  $7.4$E$+00$ & 
          $4.4$E$+01$ & $439$  \\[-0.7ex]
           &  3   &  $4.9$E$+01$ &  $6.9$E$+00$ & 
          $5.1$E$+01$ & $541$  \\[0.8ex]
     &  1   & $1.4$E$+01$ &  $2.8$E$+00$ & 
          $1.2$E$+01$ & $1200$  \\[-0.7ex]
Ly$\delta$ &  2   & $1.4$E$+01$ &  $3.6$E$+00$ & 
          $8.9$E$+00$ & $259$  \\[-0.7ex]
         &  3   &  $1.5$E$+01$ &  $2.9$E$+00$ & 
          $1.4$E$+01$ & $461$  \\[0.8ex]
     &  1   &  $1.7$E$+04$ &   $1.4$E$+03$ & 
          $1.6$E$+04$ & $25632$ \\[-0.7ex]
 \ion{Mg}{ii} k &  2   &          $1.8$E$+04$ & $1.4$E$+03$  & 
          $1.7$E$+04$ & $21165$   \\[-0.7ex]
                &  3   & $1.8$E$+04$ & $1.4$E$+03$   & 
          $1.7$E$+04$ & $14447$   \\[0.8ex]
     &  1   &  $1.3$E$+04$ &   $1.0$E$+03$ & 
          $1.3$E$+04$ & $25632$  \\[-0.7ex]
\ion{Mg}{ii} h &  2   &  $1.3$E$+04$ &   $1.0$E$+03$ & 
          $1.3$E$+04$ & $21198$    \\[-0.7ex]
      &  3   &  $1.3$E$+04$ &  $1.0$E$+03$ &
          $1.3$E$+04$ & $11813$    \\
\hline
\end{tabular}
\tablefoot{
The following histogram properties are shown in the third and further table columns:
median,\hspace{0.6ex}median of deviations from the histogram\\
\phantom{xxxxx}\hspace{1.8ex}median, the value of $E$ at which histogram
maximum occurs, and number of profiles $N$ taken into statistics.
}
\end{table}
%
%
\begin{table}[H]
\centering
\caption{Properties of histograms of $r_{\mathrm{CP}}$ shown in Fig.~\ref{Fig:a2C2P} in Appendix A for the six lines and three cases: 1\,--\,2-peak profiles from the group
2p$_{\rm ab}$, 2\,--\,from the group 2p$_{\rm a}$\,\&\,mp$_{\rm b}$, and 3\,-- combination of these two groups.}
\label{tab:appb:RCPhistoProperties}
\begin{tabular}{cccccrr}
\hline
\hline
   &      &        &              &           &            &           \\[-2.2ex]
line        &  case\hspace{3ex} & median & med. of dev. & hist. max & $N$(1-peak)  & $N$(2-peak) \\
\hline
     &      &        &              &           &            &           \\[-2.2ex]         
     &  1   &         $0.52$ &  $0.07$ & 
          $0.51$ &  $0$\hspace{3ex} & $905$\hspace{3ex} \\[-0.7ex]
 Ly$\alpha$  &  2  &         $0.51$ &  $0.05$ & 
          $0.51$ & $717$\hspace{3ex} & $1272$\hspace{3ex}  \\[-0.7ex]
      &  3  &       $0.52$ &  $0.06$ & 
        $0.49$ & $578$\hspace{3ex} & $1448$\hspace{3ex}    \\[0.8ex]
     &  1   &       $0.51$ &  $0.09$ & 
          $0.51$ & $1$\hspace{3ex} & $502$\hspace{3ex}      \\[-0.7ex]
Ly$\beta$ &  2   &       $0.38$ &  $0.05$ & 
          $0.38$ & $476$\hspace{3ex} & $216$\hspace{3ex}    \\[-0.7ex]
    &  3   &     $0.49$ &  $0.09$ & 
          $0.38$ & $260$\hspace{3ex} & $569$\hspace{3ex}  \\[0.8ex]
     &  1   &      $0.47$ &  $0.10$ & 
        $0.33$ & $2$\hspace{3ex} & $436$\hspace{3ex}  \\[-0.7ex]
Ly$\gamma$  &  2   &    $0.33$ &  $0.06$ & 
          $0.35$ & $396$\hspace{3ex} & $144$\hspace{3ex}  \\[-0.7ex]
     &  3   &  $0.45$ & $0.10$ & 
          $0.35$ & $232$\hspace{3ex} & $485$\hspace{3ex}   \\[0.8ex]
     &  1   &   $0.36$ &  $0.11$ & 
          $0.24$ & $8$\hspace{3ex} & $250$\hspace{3ex}    \\[-0.7ex]
Ly$\delta$  &  2   &  $0.24$ &  $0.08$ & 
          $0.25$ & $361$\hspace{3ex} &  $99$\hspace{3ex}  \\[-0.7ex]
            &  3   &  $0.34$ &  $0.11$ & 
          $0.24$ & $279$\hspace{3ex} & $303$\hspace{3ex}  \\[0.8ex]
     &  1   &    $0.89$ &  $0.05$ & 
          $0.93$ & $2971$\hspace{3ex} & $18193$\hspace{3ex} \\[-0.7ex]
\ion{Mg}{ii} k &  2   &   $0.75$ &  $0.04$ & 
          $0.76$ & $11765$\hspace{3ex} & $2681$\hspace{3ex}  \\[-0.7ex]
    &  3   &   $0.89$ &   $0.05$ & 
          $0.93$ &  $3911$\hspace{3ex} & $18314$\hspace{3ex}  \\[0.8ex]
     &  1   &  $0.91$ &  $0.04$ & 
          $0.94$ & $5072$\hspace{3ex} & $16125$\hspace{3ex}   \\[-0.7ex]
\ion{Mg}{ii} h &  2   &  $0.74$ &  $0.03$ &  
          $0.75$ & $10958$\hspace{3ex} & $854$\hspace{3ex}    \\[-0.7ex]
     &  3   &  $0.91$ &  $0.05$ & $0.94$ & $5789$\hspace{3ex} & $16172$\hspace{3ex}  \\
\hline
\end{tabular}
\tablefoot{Following properties are shown in the third and further table columns:
median, median of deviations from the histogram median, value of \\
\phantom{xxxxx}\hspace{1.4ex}$r_{\mathrm{CP}}$ at
which histogram maximum occurs, and numbers of
profiles assumed as 1\discretionary{-}{-}{-}peak
and 2\discretionary{-}{-}{-}peak ones in each of the three cases.
}
\end{table}
%
%
%
\begin{table}[H]
\caption{Properties of histograms of $r_{\mathrm{PA}}$ shown in Fig.~\ref{Fig:a3PAs}
         for the six lines and three cases: 1\,--\,2-peak profiles from the group
         2p$_{\rm ab}$, 2\,--\,from the group 2p$_{\rm a}$\,\&\,mp$_{\rm b}$,
         and 3\,--\,combination of these two groups.}
\label{tab:appb:RPAhistoProperties}
\centering
\begin{tabular}{cccccccccrr}
\hline
\hline
line &  case  & \multicolumn{3}{c}{$r_{\mathrm{B-R}}$} &  &
    \multicolumn{3}{c}{$r_{\mathrm{R-B}}$}  &   
$N$(2-peak sm)    &  $N$(2-peak asm)  \\
    \cline{3-5} \cline{7-9}
   &     & median & med. of dev. & hist. max &  &
median & med. of dev. & hist. max  & 
      &     \\
\hline
     &      &      &    &     &     &      &      &     &   &  \\[-2.3ex]
     &  1   &  $0.89$ &  $0.05$ &  $0.91$ &   & $0.86$ &  
  $0.06$ &  $0.93$  
 &   $64$\hspace{6ex} &  $841$\hspace{6ex} \\[-0.7ex]
  Ly$\alpha$ &  2   & $0.91$ &  $0.05$ &  $0.93$ &   & $0.88$ &  
  $0.07$ &  $0.93$  & $129$\hspace{6ex} & $1143$\hspace{6ex} \\[-0.7ex] 
    &  3   &  $0.90$ & $0.05$  & $0.93$  &  & $0.87$ &  
    $0.06$ & $0.93$  &  $134$\hspace{6ex} & $1314$\hspace{6ex}  \\[0.8ex]
     &  1   & $0.89$ & $0.06$  & $0.95$ &  & $0.78$ & $0.09$ & $0.69$  & $34$\hspace{6ex} &  
     $468$\hspace{6ex}  \\[-0.7ex]
Ly$\beta$   &  2   &  $0.91$ & $0.05$  & $0.90$ &  & $0.87$ & $0.08$ & $0.88$  
 & $25$\hspace{6ex} &  $191$\hspace{6ex}  \\[-0.7ex]
   &  3   &  $0.90$ &  $0.05$  & $0.90$ &  & $0.79$ & $0.08$ & $0.69$ & $38$\hspace{6ex} & 
   $532$\hspace{6ex}    \\[0.8ex]
     &  1   &  $0.87$ &  $0.07$ & $0.96$ &  & $0.77$ &  $0.11$ & $0.64$  & $42$\hspace{6ex} 
     & $394$\hspace{6ex} \\[-0.7ex]
Ly$\gamma$  &  2   & $0.89$ &  $0.06$ & $0.96$ &  & $0.87$ & $0.07$ & $0.96$  
 & $28$\hspace{6ex} & $116$\hspace{6ex}    \\[-0.7ex]
  &  3   &  $0.87$ & $0.07$  & $0.96$ &  & $0.79$ &  $0.11$ & $0.89$  & 
  $52$\hspace{6ex} & $433$\hspace{6ex}     \\[0.8ex]
     &  1   & $0.83$ & $0.10$ & $0.80$ &  &  $0.75$ &  $0.13$ & $0.70$  
 & $24$\hspace{6ex} & $226$\hspace{6ex}    \\[-0.7ex]
Ly$\delta$  &  2   & $0.78$ & $0.20$ & $0.95$ &  & $0.78$ & $0.09$ & $0.70$ & 
 $14$\hspace{6ex} & $85$\hspace{6ex}  \\[-0.7ex]
   &  3   & $0.82$ &   $0.11$  & $0.85$  &  & $0.75$ & $0.11$ & $0.70$ & $31$\hspace{6ex} 
    & 272\hspace{6ex} \\[0.8ex]
     &  1   & $0.91$ & $0.05$ & $0.96$ &  & $0.93$ & $0.05$ & $0.98$ & $2412$\hspace{6ex} 
     & $15781$\hspace{6ex}  \\[-0.7ex]
\ion{Mg}{ii} k &  2  & $0.91$ & $0.05$  & $0.98$ &  & $0.940$ & $0.04$ & $0.98$  
 &  $361$\hspace{6ex} & $2320$\hspace{6ex}  \\[-0.7ex]
   &  3  & $0.91$ & $0.05$ & $0.96$ &  & $0.93$ & $0.05$ & $0.98$ & $2447$\hspace{6ex} 
   & $15867$\hspace{6ex}    \\[0.8ex]
     &  1   & $0.93$ &  $0.04$ & $0.98$ &  & $0.94$ & $0.04$ & $0.98$   
 & $2677$\hspace{6ex} & $13448$\hspace{6ex}   \\[-0.7ex]
\ion{Mg}{ii} h & 2  & $0.92$ & $0.06$ & $0.98$ &  & $0.95$ & $0.03$ & $0.98$  
 &  $164$\hspace{6ex} & $690$\hspace{6ex}    \\[-0.7ex]
  & 3  & $0.93$ & $0.04$ & $0.98$ &  & $0.94$ & $0.04$ & $0.98$ & $2683$\hspace{6ex} & 
  $13489$\hspace{6ex} \\[0.2ex]
\hline
\end{tabular}
\tablefoot{
Following properties are shown in the third and further table columns:
median, median of deviations from the
  histogram median, value of\\
  \phantom{xxxxx}\hspace{1.4ex}$r_{\mathrm{PA}}$ at which histogram maximum
  occurs -- for both blue\discretionary{-}{-}{-}to\discretionary{-}{-}{-}RED and
  red\discretionary{-}{-}{-}to\discretionary{-}{-}{-}BLUE histograms
  ($r_{\mathrm{B-R}}$ and $r_{\mathrm{R-B}}$).\hspace{1.2ex} In the last two
  columns,\\
  \phantom{xxxxx}\hspace{1.4ex}there are shown numbers of profiles with symmetrical
  and asymmetrical peaks within those assumed as
  2\discretionary{-}{-}{-}peak ones in the three cases.}
\end{table}
%
%
\addtolength{\topmargin}{-0.3cm}
\addtolength{\topskip}{-0.5cm}
\end{appendix}
\end{document}